\documentclass[a4paper,times,fleqn]{cas-dc}
\usepackage[numbers]{natbib}

\usepackage{graphicx}\graphicspath{{images/}}
\usepackage{enumitem}
\usepackage{multicol}
\usepackage{multirow}
\usepackage{longtable}
\usepackage{lscape}
\usepackage{float}
\usepackage{tablefootnote}
\usepackage{tabularx}
\usepackage{booktabs}
\usepackage{adjustbox}
\usepackage{flushend}

\begin{document}
\let\WriteBookmarks\relax
\def\floatpagepagefraction{1}
\def\textpagefraction{.001}

\shorttitle{Cybersecurity in V2G Systems}
\shortauthors{M A Razzaque et~al.}

\title [mode = title]{Cybersecurity in Vehicle-to-Grid (V2G) Systems: A Systematic Review}                      



%

\author[1]{Mohammad A. Razzaque}[orcid=0000-0002-5572-057X]

\ead{m.razzaque@tees.ac.uk}

\author[2]{Shafiuzzaman K Khadem} [orcid=0000-0001-5869-770X]
\ead{shafi.khadem@ierc.ie}

\author[2]{Sandipan Patra} [orcid=0000-0001-6018-6984]
\ead{sandipan.patra@ierc.ie}

\author[1]{Glory, Okwata} []
\ead{okwataglory2050@gmail.com}

\author[3]{Md. Noor-A-Rahim} [orcid=0000-0003-0587-3145]
\ead{m.rahim@cs.ucc.ie}



\affiliation[1]{organization={School of Computing, Engineering and Digital Technologies, Teesside University}, country={UK}}

\affiliation[2]{organization={International Energy Research Centre, Tyndall National Institute, Cork}, country = {IE}}

\affiliation[3]{organization={nasc Research, School of Computer Science \& IT, University College Cork}, country = {IE}}

\cortext[cor1]{Corresponding author: Mohammad A. Razzaque}



\begin{abstract}
Integrating electric vehicles (EVs) into the power grid through vehicle-to-grid (V2G) systems offers transformative potential for energy optimisation and grid stability. However, this bidirectional energy exchange introduces significant cybersecurity challenges, including vulnerabilities to spoofing, denial-of-service attacks, and data manipulation, which threaten the integrity and reliability of the V2G system. Despite the growing body of research on V2G cybersecurity, existing studies often adopt fragmented approaches, leaving gaps in addressing the entire ecosystem, including users, electric vehicles, charging stations, and energy market and trading platforms. This paper presents a systematic review of recent advancements in V2G cybersecurity, employing the PRISMA (Preferred Reporting Items for Systematic Reviews and Meta-Analyses) framework for detailed searches across three journal databases and included only peer-reviewed studies published between 2020 and 2024 (June). We identified and reviewed 133 V2G cybersecurity studies and found five important insights on existing V2G cybersecurity research. \textit{First}, most studies (103 of 133) focused on protecting V2G
systems against cyber threats, while only seven studies addressed the recovery aspect of the CRML (Cybersecurity Risk Management Lifecycle) function. \textit{Second}, existing studies have adequately addressed the security of EVs and EVCS (EV charging stations) in V2G systems (112 and 81 of 133 studies, respectively). However, none have focused on the linkage between the behaviour of EV users and the cybersecurity of V2G systems. \textit{Third}, physical access, control-related vulnerabilities, and user behaviour-related attacks in V2G systems are not addressed significantly. Furthermore, existing studies overlook vulnerabilities and attacks specific to AI and blockchain technologies. \textit{Fourth}, blockchain, artificial intelligence (AI), encryption, control theory, and optimisation are the main technologies used, and \textit{finally}, the inclusion of quantum safety within encryption and AI models and AI assurance (AIA) is in a very early stage; only two and one of 133 studies explicitly addressed quantum safety and AIA through explainability. By providing a holistic perspective, this study identifies critical research gaps and outlines future directions for developing robust end-to-end cybersecurity solutions to safeguard V2G systems and support global sustainability goals.

\end{abstract}



\begin{keywords}

Elective Vehicles (EV) \sep Smart Grid \sep Autonomous Vehicles \sep Cybersecurity \sep Blockchain \sep Artificial Intelligence \sep Vulnerabilities \sep Attacks \sep Quantum-safe \sep AI Assurance 

\end{keywords}

\maketitle

\section{Introduction}
\label{sec1}

Electric vehicles (EVs) are emerging as a cornerstone of sustainable transportation, offering multifaceted benefits such as significantly reducing CO2 emissions - 50g/mile compared to 225g/mile for gasoline vehicles by 2050~\cite{Areelect30:online}. This environmental advantage, coupled with economic incentives, is driving a global shift toward electric mobility. For example, EV sales are projected to increase from 13.68 million units in 2024 to 18.84 million units by 2029~\cite{Electric33:online}, with EV sales expected to surpass those of gas vehicles by 2040~\cite{ReportEV3:online}. Rapid adoption of EVs marks a critical milestone in reducing greenhouse gas emissions and advancing global sustainability goals.

As EV adoption proliferates, their integration with the power grid through vehicle-to-grid (V2G) systems presents transformative opportunities for energy optimisation and grid stability. V2G systems enable bidirectional energy flow, allowing EVs to function as mobile energy storage units. These systems can supply power to the grid during peak demand periods while facilitating the integration of renewable energy sources~\cite{electronics13101940}. The practicality of such systems is underscored by the fact that vehicles remain idle up to 95\% of the time~\cite{Electric27:online}, making bidirectional integration a logical extension of their utility.

Despite these advantages, V2G systems face significant challenges~\cite{aljohani2024comprehensive, acharya2020cybersecurity, ronanki2023electric}. These include interoperability, data privacy, and the increasing risk of cybersecurity threats in an increasingly digitised and bidirectionally connected ecosystem. Cyberattacks targeting various components of the V2G system - including EVs, EV charging stations (EVCS), the utility grid, users, energy market and trading platforms, and third-party entities - pose serious risks. These attacks can cause power outages, data breaches, and system manipulation, compromising the integrity and availability of critical infrastructure. For example, communication between EVCS and the grid is vulnerable to threats such as spoofing, denial of service (DoS) attacks, and data manipulation~\cite{ronanki2023electric, acharya2020cybersecurity, Paper-6}. A DC link short circuit attack on an EVCS could significantly damage power converters and even disrupt grid operations~\cite{ronanki2023electric}. Addressing these vulnerabilities is critical to ensure the secure and reliable operation of V2G systems.


Realising the importance of cybersecurity of V2G systems, many research works have been published, including several recent ones~\cite{Paper-2, Paper-3, Paper-4, Paper-5, Paper-6, Paper-7, Paper-8, Paper-10, Paper-12}. These works are highly diverse, mainly in terms of (i) the Cybersecurity Risk Management Lifecycle's (CRML) functions, (ii) vulnerabilities and attack vectors, and (iii) addressed cybersecurity principles (e.g., CIA triad), (iv) considered elements of V2G, (v) key technologies (e.g., Blockchain, AI) in solving cybersecurity issues, and (vi) the quality of the provided solution (e.g., AI assurance, quantum-safe encryption). For example, ring signatures, encryption, and certificate-less cryptography were used~\cite{Paper-2} to protect the grid system from data falsification and impersonation attacks. The EV charger-based voltage control scheme was used in~\cite{Paper-3} to protect EVCS and the grid from measurement value falsification and FDIA (false data injection attack), which generally have fragile communication and physical access to EVs (vulnerabilities). The authors in~\cite{Paper-5} used hybrid deep learning (DL) and explainable AI (XAI) to detect malware injection (attack) in EV and EVCS. On the other hand, the authors in~\cite{Paper-64} investigate the security posture of mobile EV charging applications as an attack surface in the EV charging ecosystem and the underlying infrastructure using different security analysis tools. 

Considering the significant potential of V2G systems, such as reducing $\text{CO}_2$ emissions and improving grid stability, and their essential role as a critical national infrastructure, robust end-to-end cybersecurity is necessary. This encompasses protection across the entire V2G ecosystem, including users, EVs, EVCS, energy market and trading platforms, third-party stakeholders, and the grid. Achieving such comprehensive security requires a complete review of existing cybersecurity solutions for V2G systems and adopting a holistic perspective to identify research gaps and areas that require further investigation. Several review papers~\cite{hou2024cyber, aljohani2024comprehensive, ronanki2023electric, achaal2024study, hamdare2023cybersecurity,acharya2020cybersecurity} have explored different dimensions of V2G cybersecurity. As highlighted in \cite{hou2024cyber},  the study examines the cyber resilience of power electronics-enabled systems, particularly renewable energy sources (RES), against sophisticated attacks such as DoS and FDIA. Another comprehensive survey~\cite{aljohani2024comprehensive} investigates cyber-physical vulnerabilities specific to electric vehicles, including DoS, jamming, man-in-the-middle (MITM) and spoofing attacks. These vulnerabilities often arise from insecure controller area networks (CAN), battery management systems (BMS), and weak authentication mechanisms. Similarly, Chandwani et al.~\cite{chandwani2020cybersecurity} address cybersecurity issues related to EV onboard charging (OBC) systems, focusing on control-based threats such as CAN-bus manipulation and intra-FPGA attacks, as well as hardware-based risks such as sudden load loss and grid side short circuits. However, many of these studies take a fragmented view of V2G security. For example, while Acharya et al.~\cite{acharya2020cybersecurity} analyse cybersecurity concerns in EVs, EVCS, and the grid, they neglect critical aspects such as user interactions and energy trading.

Furthermore, these reviews do not cover recent advancements in the field. To address these gaps, it is essential to conduct an integrated and up-to-date assessment of cybersecurity challenges and solutions that include all components of the V2G ecosystem. This study aims to bridge these gaps by providing a holistic perspective through a systematic review of recent research on V2G cybersecurity. Specifically, the key features of these studies are identified and an in-depth analysis of those features is provided, highlighting critical advancements, limitations, and areas for further exploration. This endeavour consolidates current knowledge and facilitates the advancement of more effective cybersecurity solutions for V2G or relevant application domains. The primary contributions of this study are as follows.
\begin{itemize}
    \item For a better systematic review and understanding of cybersecurity issues, the V2G system is presented as a complex Cyber-Physical System (CPS).

    \item A comprehensive and systematic review of existing V2G cybersecurity studies using the outlined features and the PRISMA (Preferred Reporting Items for Systematic Reviews and Meta-Analyses) framework-based systematic selection approach of existing studies.
        \item Identification of key features and overview of recent research on V2G cybersecurity. Most of these are system-agnostic and can be used in any cybersecurity study and 
    \item An overview of the open research challenges and future directions in end-to-end cybersecurity of V2G systems.
    
\end{itemize}

The remainder of the paper is structured as follows. Section~\ref{sec:2} provides an overview of the V2G system from the perspective of CPS and reviews the existing literature and related studies on V2G cybersecurity. Section~\ref{sec3} deals with the key features of recent V2G cybersecurity research. Section~\ref{sec:3b} outlines the methodology for investigating and selecting relevant V2G cybersecurity studies. The results of these studies, categorised by identified key features, are presented in Section~\ref{sec:4}. Section~\ref{sec6} highlights open research challenges and proposes future research directions. Finally, Section~\ref{sec:6} concludes the paper and discusses potential avenues for future work.

\begin{figure*}[hbt!]
\centering
\includegraphics[width=17.5cm]{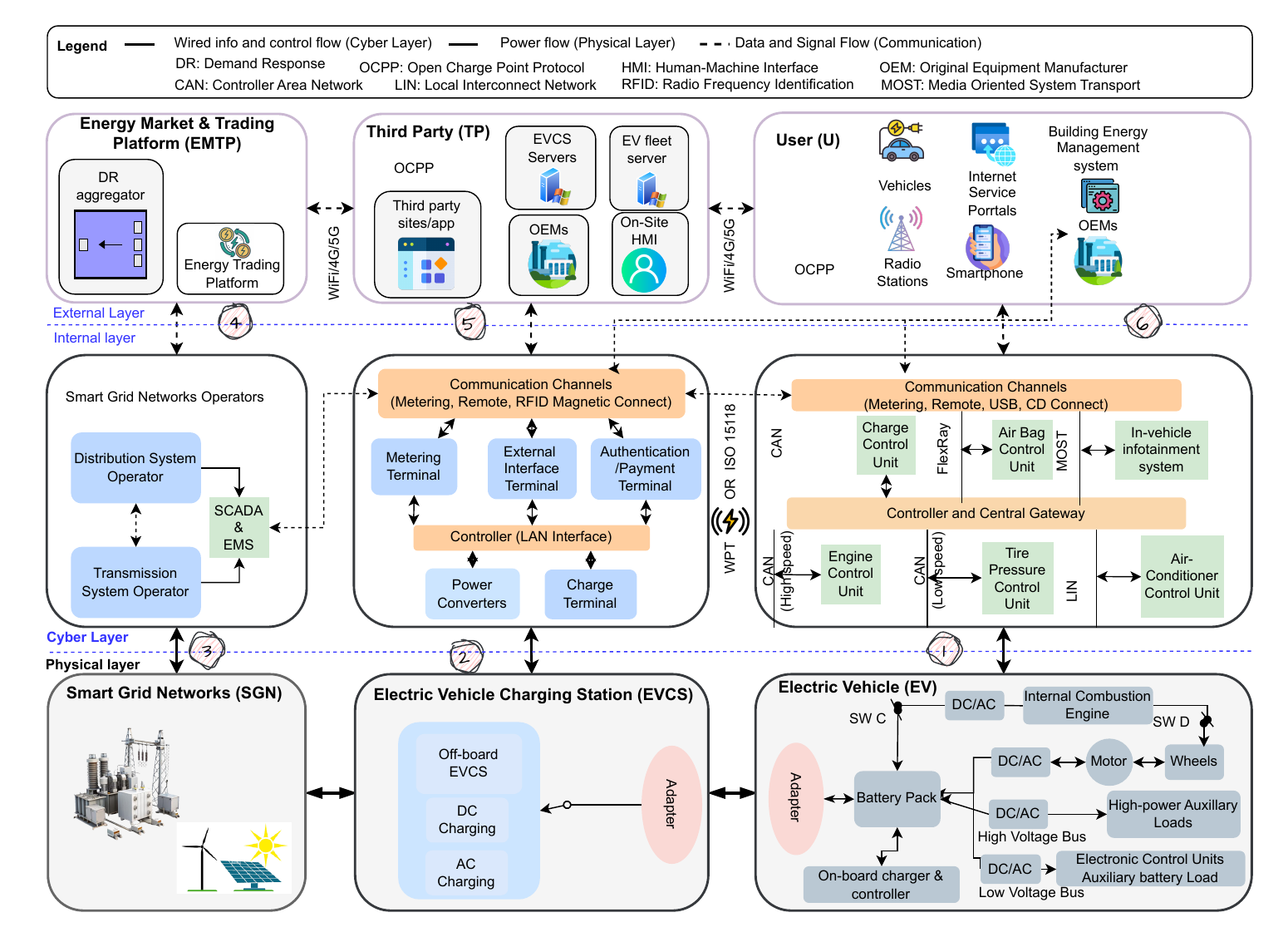}
\caption{A multi-level and detailed view of a V2G system.}
\label{fig1}
\end{figure*}



\section{Background and Related Works}
\label{sec:2}
\subsection{Cyber-Physical System (CPS) Overview}
A V2G system is a complex CPS that enables bidirectional power flow between electric vehicles and the power grid. The system comprises six key components (shown in Figure~\ref{fig1}) that work together to facilitate the functionality of V2G~\cite{acharya2020cybersecurity, chandwani2020cybersecurity,achaal2024study}.

\begin{enumerate}
    \item Electric Vehicles (EVs): EVs serve as mobile energy storage units in the V2G ecosystem. They consist of physical components such as motors, charging cables, battery, steering, wheels and onboard chargers, as well as cyber elements (internal cyber layer in Figure~\ref{fig1}) such as Electronic Control Units (ECUs), battery management system (BMS), On-Board Diagnostics ports, communication systems (e.g., CAN bus) and firmware installed within the ECU or other devices.
    
    \item EVCSs: Charging stations facilitate energy transfer between vehicles and the grid. They include physical infrastructure such as off-board chargers and controllers and cyber components (internal cyber layer in Figure~\ref{fig1}) such as payment terminals, power converters, RFID card reader, and controller LAN (local area network), firmware or software, protocols (e.g., OCCP, OpenADR) to communicate with external cyber layer components, such as EVCS server or Building Energy Management Systems.
    
    \item Smart Grid Networks (SGN): The power grid, supplemented by distributed energy resources (DERs) such as solar panels and wind turbines, ensures energy generation, distribution, and storage. This component integrates the physical infrastructure with cyber elements, such as SCADA (supervisory Control and Data Acquisition), including sensors (e.g., RTU-remote terminal unit, PMU-phasor measurement unit),  control systems (e.g., programmable logic controller), and communication protocols (e.g., OCCP, OpenADR) to communicate with external cyber layer components, such as demand response aggregator.
    
    \item Energy Market and Trading Platforms (EMTP): EMTPs provide a marketplace for dynamic energy transactions, enabling stakeholders to optimise costs and energy utilisation. These platforms operate primarily on the cyber layer (external cyber layer in Figure~\ref{fig1}), facilitating data exchange and coordination throughout the V2G ecosystem.
    
    \item Third-party Entities (TP): Original equipment manufacturers, road side units, mobile apps for charging EVs, on-site human-machine interface, and other third-party entities contribute to the V2G system through hardware innovation, software updates, and support services. They are crucial in enhancing the system's capabilities and ensuring smooth operation.
    
    \item End-users (U): Individual and institutional users participate as energy consumers, producers, or both. They shape the system's demand-response dynamics and interact with the V2G infrastructure through various interfaces, such as smartphones and charging station controls.

\end{enumerate}

Understanding the CPS aspect of the V2G systems, as illustrated in Figure~\ref{fig1}, is essential to recognise and assess cybersecurity risks and threats. Some of the important reasons are the following.
\begin{itemize}
    \item Integration of Physical and Cyber Layers: The cyber-physical nature of V2G systems means that physical components, such as electric vehicles, electric vehicle charging stations, and the power grid, are tightly integrated with cyber components, including communication networks, control systems, and data management platforms. This integration enables advanced functions such as bidirectional power flow and real-time monitoring. However, it also creates vulnerabilities, as attacks on the cyber layer can cascade into physical disruptions, compromising grid stability or damaging infrastructure~\cite {aljohani2024comprehensive, diaba2024cyber,chandwani2020cybersecurity,achaal2024study}.

    \item Real-Time Monitoring and Response: Cyber-physical systems rely on real-time data exchanges for effective operation. Sensors monitor various parameters, such as energy flow, grid frequency, and battery status, while control algorithms manage the charging and discharging process. Cyberattacks, such as false data injection, can distort these metrics, leading to incorrect decisions within control systems. A robust understanding of the CPS framework allows the development of anomaly detection systems that can quickly identify and mitigate such attacks~\cite {aljohani2024comprehensive, diaba2024cyber,chandwani2020cybersecurity,achaal2024study}.

    \item Increased Attack Surface: In V2G systems, multiple stakeholders, such as EV owners, grid operators, charging infrastructure providers, and third-party platforms, interact through interconnected systems. Each connection introduces potential entry points for cyberattacks. For example, a compromised EVCS could be a launch point for attacks on the larger network and the energy trading mechanism. Understanding the aspect of CPS helps identify such vulnerabilities and develop tailored security measures to protect the entire ecosystem~\cite {aljohani2024comprehensive, diaba2024cyber,chandwani2020cybersecurity,achaal2024study}.
    
    \item Threats to Critical Infrastructure: V2G systems are integral to smart grids and are classified as critical infrastructure. Disruptions caused by cyberattacks, such as Denial-of-Service (DoS) attacks, spoofing, or ransomware, can lead to widespread power outages, economic losses, and safety risks. For example, a successful attack on an EV's charging protocol could manipulate energy transfer, overload the grid, or interrupt the power supply to essential services~\cite{aljohani2024comprehensive,Darktrac24}. Researchers and engineers can prioritise securing critical nodes and communication pathways by understanding CPS interactions.

    \item Evolving Threat Landscape: As V2G technology evolves, the complexity of threats also increases. A robust understanding of cyberphysical interactions helps to anticipate emerging vulnerabilities and design adaptive security measures~\cite{diaba2024cyber, achaal2024study}.
    \item Security by Design: Addressing cybersecurity risks in V2G systems requires integrating security measures into the design phase of both physical and cyber components. By understanding the interdependencies of the CPS, stakeholders can adopt a "security by design" approach, incorporating features such as secure communication protocols, authentication mechanisms, and hardware-based protections directly into system architectures~\cite{moghadasi2022trust}.
   
    \item Holistic security approach: Recognising the cyber-physical aspect encourages a comprehensive security strategy that addresses digital and physical vulnerabilities rather than treating them in isolation~\cite{diaba2024cyber,chandwani2020cybersecurity,achaal2024study}.

\item Facilitating Regulatory Compliance: Many regions have introduced standards and regulations to secure CPSs, particularly in critical infrastructure sectors. Understanding the CPS structure is essential to ensure compliance with these standards. For example, protocols such as ISO 15118 for secure EV charging require encryption and authentication measures, which can only be effectively implemented through a clear understanding of the dynamics of CPS~\cite{ISO151187:online}.
\end{itemize}

\begin{figure*}[hbt!]
\centering
\includegraphics[width=17.0cm]{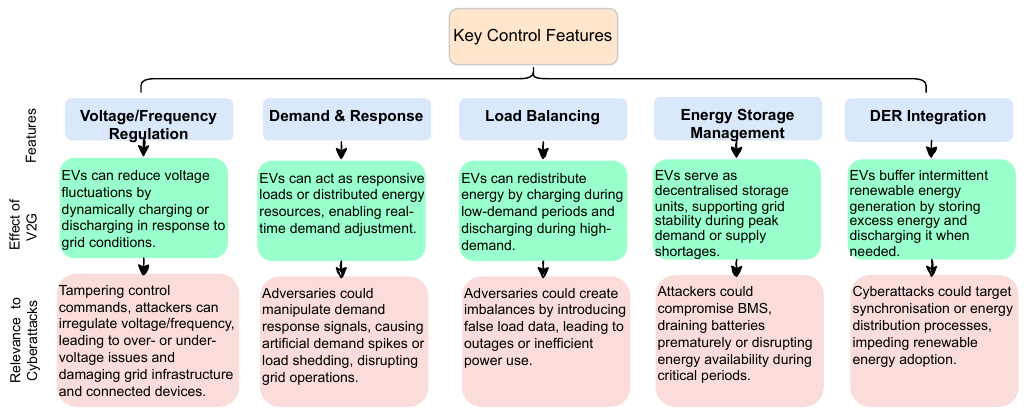}
\caption{Key control features, effects of V2G and relevance to cyberattacks on power grid.}
\label{fig2a}
\end{figure*} 

\subsection{Effects of V2G on SGN and Relevance to Cyberattacks}

As illustrated in Figure~\ref{fig1}, a V2G system consists of at least three key cyber-physical systems: EV, EVCS and the power grid. Bidirectional communication and control of these CPSs through V2G systems introduce both positive and negative effects~\cite{MASTOI20231777, MASTOI202211504}, particularly on EVs and the grid. For example, EV owners can discharge energy to support the grid, while the grid benefits from load balancing made possible by the contributions of EVs.

This bidirectional connectivity supports essential control functions, including frequency and voltage regulation, demand response, load balance, energy storage management, and DER integration. However, it also introduces vulnerabilities that cybercriminals can exploit, as summarised in Figure~\ref{fig2a}. These vulnerabilities, such as insecure communication channels, can significantly increase the risk of cyberattacks on the SGN, adversely affecting these critical control features. For example, cyber-physical attacks that leverage EVCS botnets are a notable threat. Such attacks can introduce generator oscillations~\cite{Paper-6}, destabilising the SGN. Similarly, malicious mode attacks (MMAs)—a recent cyberattack pattern—generate high-amplitude forced oscillations, severely threatening the stability of SGN and disrupting power supplies to charging stations. Furthermore, attacks involving demand falsification or manipulation can compromise the demand-response equilibrium of V2G systems, increasing instability and reducing system efficiency~\cite{Paper-52}.

\begin{table*}[]
\caption{Related study: Existing and related review or survey papers on Cybersecurity}
\label{tab2}
\renewcommand{\arraystretch}{1.2} 
\resizebox{\textwidth}{!}{%
\begin{tabular}{|l|l|l|l|l|l|l|}
\hline
\textbf{Study} & \textbf{V2G Elements} & \textbf{Key Vulnerabilities} & \textbf{ Main AVs/A} & \textbf{SR} & \textbf{Limitations} & \textbf{PY} \\ \hline
\cite{hou2024cyber}         & Grid                    & RES integration related                     & PE-related (DoS, FDIA, RA, MI)      & No & System-level analysis                & 2024 \\ \hline
\cite{diaba2024cyber} & Grid & WAu, IC & PT, UA, DoS, MITM & No &  Not on V2G, and limited attack analysis & 2024\\ \hline

\cite{aljohani2024comprehensive} & EV       & Insecure CAN bus, BMS, CCo                  & DoS, MITM, Jamming, Spoofing        & No & U, EMTP, and Grid side missing        & 2024 \\ \hline
\cite{ronanki2023electric}  & EV, EVCS, and Grid       & IC, INCSs, WAu    & DM, IS/AS, IC, UA  & No & U, EMTP, and Grid side missing        & 2024 \\ \hline
\cite{achaal2024study}      & Grid             & Insecure AGC, SCADA, and IC & MITM, FDIA, TSA, DoS, MI      & No & Not on V2G, U missing               & 2024 \\ \hline

\cite{hamdare2023cybersecurity} & EV and EVCS            & IC (e.g., ISO 15118) and HV              & MITM, DB, UA, PT                    & No & Vulnerabilities \& AV/A not comprehensive & 2023 \\ \hline
\cite{zografopoulos2023distributed} & EV, EVCS and Grid & Insecure Modbus, DNP3, OpenADR, IoT, PA     & LoI, DM, SCA, IF                   & No & Mostly DER-related                   & 2023 \\ \hline
\cite{hasan2022blockchain}  &  EMTP and SGN                & Insecure AGC, SCADA, and IC                & MITM, FDIA, Jamming, DM            & No & Not on V2G, EMTP missing               & 2022 \\ \hline
\cite{zografopoulos2021cyber} & Grid                  & Insecure PMU and GPS, and PA              & DM, TDA, GPS spoofing                    & No & Not on V2G systems                   & 2021 \\ \hline
\cite{miglani2020blockchain} & Grid & IC  & Malware, DDoS, Jamming, Phishing    & No & Not on V2G systems    & 2020 \\ \hline
\cite{chandwani2020cybersecurity} & EV & Insecure CAN bus, BMS, CCo                  & DM, IC, UA & No & Only OBC-related, recent studies missing & 2020 \\ \hline
\cite{acharya2020cybersecurity} & EV, EVCS, and Grid    & Insecure PMU, RUTs, CAN bus  & Spoofing, MITM, RA, Tampering       & No & U and EMTP missing      & 2020 \\ \hline
\cite{hossain2020cyber}     & Grid    & CPS-related  & DoS, FDIA, MITM, PAs                & No & Missing recent studies               & 2019 \\ \hline
\multicolumn{7}{@{}|p{25cm}|}{\scriptsize \textbf{Acronyms:} AGC (Automatic Generation Control), AVs/A (Attack Vectors/Attacks), BMS (Battery Management System), CAN (Controller Area Network), CCo (Charger Controller), DB (Data Breaches), DM (Data or Signal Modification), DoS (Denial of Service), EMTP (Energy Market and Trading Platform), EV (Electric Vehicle), EVCS (Electric Vehicle Charging Station), FDIA (False Data Injection Attack), IC (Insecure Communication), IC/S (Interruption (communication channels/sensors)), INCSs (Insecure Networked Control Systems), HV (Hardware-related vulnerabilities), MI (Malicious Injection), MITM (Man-in-the-Middle), PA (Physical Access), PT (Physical Tempering), PY (Publication Year), RA (Replay Attack), RES (Renewable Energy Sources), SC (Side Channel), SCA (Supply Chain Attacks), SCADA (Supervisory Control and Data Acquisition), SGN (Smart Grid Networks), SQLI (SQL Injection), SR (Systematic Review), TDA (Time Delay Attack), TSA (Time Synchronisation Attacks), UA (Unauthorised Access), U (Users), V2G (Vehicle-to-Grid), WAu (Weak Authentication).} \\ \hline
\end{tabular}%
}
\end{table*}

\subsection{Related Works}

Researchers have paid significant attention to cybersecurity in V2G systems due to its critical role in the future of smart energy systems. Several comprehensive reviews, including~\cite{hou2024cyber, aljohani2024comprehensive, ronanki2023electric, achaal2024study, hamdare2023cybersecurity,acharya2020cybersecurity} have addressed various aspects of this domain, each contributing unique insights and revealing areas that require further exploration. 


The authors in \cite{hou2024cyber} have thoroughly reviewed cyber resilience in power systems enabled by power electronics, discussing vulnerabilities and defence strategies in various components. Although this work offers valuable information on the power electronics devices and components level, it lacks a holistic system-level perspective crucial to understanding V2G cybersecurity in a comprehensive way. Similarly, cyber-physical attacks on future energy systems have been explored in \cite{diaba2024cyber}, with a focus on smart grid vulnerabilities, such as weak passwords (WP), insecure communication (IC), attacks including DoS, MITM, and their security measures. The proposed multidisciplinary framework is useful, but the paper has not explicitly reviewed the V2G system, and attack analysis is limited.


A detailed overview of EV cybersecurity, categorising threats, defences, and testing methods is presented in~\cite{aljohani2024comprehensive}. However, the paper lacks specific real-world case studies and an in-depth analysis of the effectiveness of defensive techniques, particularly in areas beyond EVs and EVCS, including user and energy trading platforms. A comprehensive review of EVCS technologies and their cybersecurity implications has been presented in \cite{ronanki2023electric}. The analysis of data-integrity-based cyberattacks on various charging systems through case studies is particularly valuable. However, the paper could benefit from investigating specific attack scenarios and quantifying potential impacts, especially with regard to the user, grid, and energy trading aspects of V2G.


The authors in \cite{achaal2024study} provide a comprehensive overview of Smart Grid (SG) cybersecurity, proposing a cyberattack taxonomy based on the US National Institute of Standards and Technology (NIST) cybersecurity framework. However,  this SG-related review is not specific to V2G, and user aspects are missing. On the other hand, \cite{zografopoulos2023distributed} focuses on the cybersecurity outlook of DER, including EVCS. Their analysis of vulnerabilities in the CPS layers of DERs and proposed mitigation strategies is helpful. However, the paper primarily considers DERs part of V2G systems and lacks detailed case studies that demonstrate the real-world impacts of cyberattacks on specific DER systems. A comprehensive overview of CPS security, focusing on cyber-physical energy systems (CPES) is presented in \cite{zografopoulos2021cyber}. Although their structured analysis is helpful for understanding CPS-related attacks, the paper lacks a specific discussion of V2G system vulnerabilities. 


Blockchain is a key solution technology for security measurements and energy trading (ET) platforms. Hasan et al. \cite{hasan2022blockchain} review blockchain implementations in smart grids, focusing on security and data protection aspects in energy trading. Although valuable, the paper lacks a dedicated analysis of how blockchain can specifically enhance V2G security. The authors in \cite{miglani2020blockchain} explore the use of blockchain to secure Internet of Energy (IoE) systems. However, the paper lacks a detailed discussion of practical challenges and scalability issues in implementing blockchain in large-scale IoE deployments, particularly in the context of V2G systems.


The authors in \cite{hamdare2023cybersecurity} explore cybersecurity risks and validate threats in EVCS networks using real-time data. Although providing valuable insights, the article lacks comprehensive vulnerability analysis and in-depth mitigation strategies. Chandwani et al. \cite{chandwani2020cybersecurity} investigated data integrity attacks targeting EV onboard chargers (OBCs), proposing software and hardware countermeasures. However, the paper's focus on OBC systems limits its applicability to the broader V2G ecosystem, and recent studies are missing. In another work, Acharya et al. \cite{acharya2020cybersecurity} present a comprehensive review of the device- and network-level vulnerabilities in electric vehicles, charging stations and their interaction with the power grid. They also present a threat model using an attack tree. However, they do not cover physical security threats to the EV charging infrastructure. In addition, cybersecurity related to users (U) and the energy market and trading (EMT) is missing in recent studies.


The existing and related reviews on cybersecurity issues of V2G are summarised in Table~\ref{tab2} in terms of (i) V2G related elements, (ii) key vulnerabilities, (iii) attack vectors or attacks, and (iv) whether the review was carried out systematically (e.g. using a framework like PRISMA) or not, and their limitations. None of the existing work systematically reviewed and considered the cybersecurity of the six elements of V2G (i.e., EV, EVCS, SGN, U, EMT platforms, and TP), especially the perspectives of cybersecurity of U or EMT platforms.

\begin{figure*}[hbt!]
\centering
\includegraphics[width=17cm]{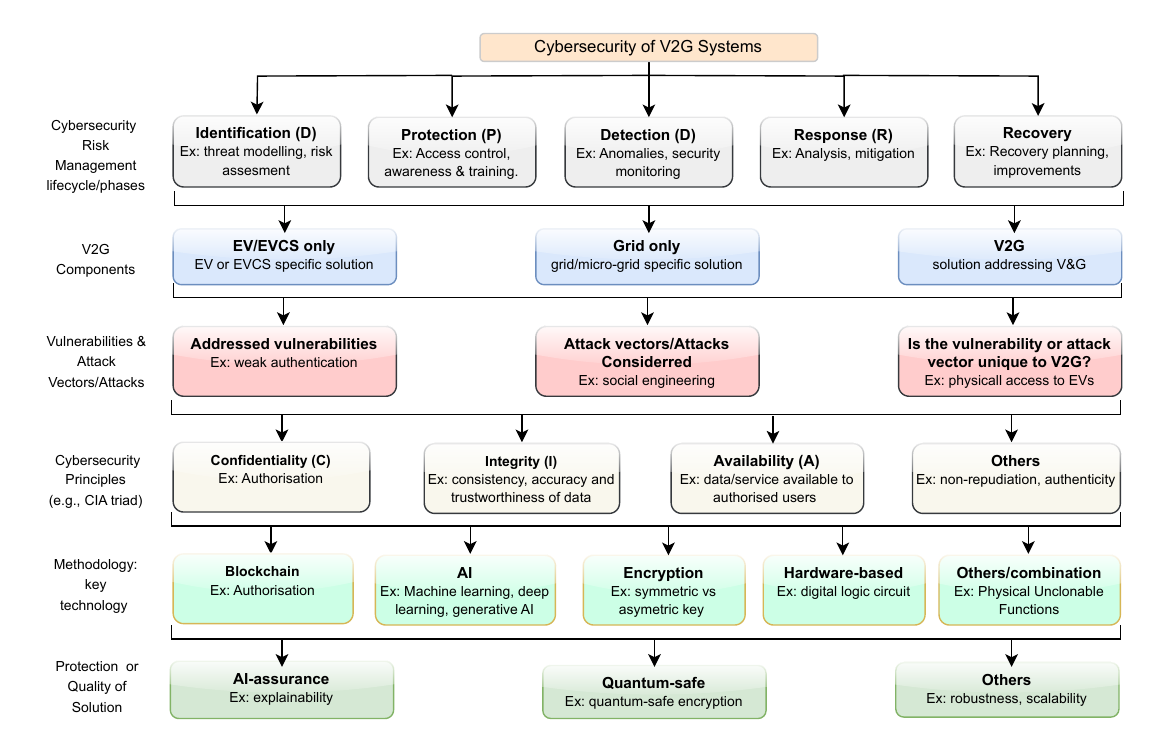}
\caption{Key features of existing V2G's cybersecurity-related works.}
\label{fig3a}
\end{figure*}

\section{Key Features of the V2G's Cybersecurity}
\label{sec3}

Many research studies have been published~\cite{Paper-2, Paper-3, Paper-4, Paper-5, Paper-6, Paper-7, Paper-8, Paper-10, Paper-12} on various aspects/features of cybersecurity (e.g., vulnerabilities, attacks, or attack vectors) and V2G systems (e.g., cyber-physical aspects and V2G elements considered). Identification of these features and review of existing cybersecurity research on V2G systems from these features' perspectives is crucial to gain a comprehensive understanding of the field, identifying the gaps and contributing to future research. Figure~\ref{fig3a} summarises these key features, and the following subsections provide an overview of those features.

\begin{figure*}[hbt!]
\centering
\includegraphics[width=17.0cm]{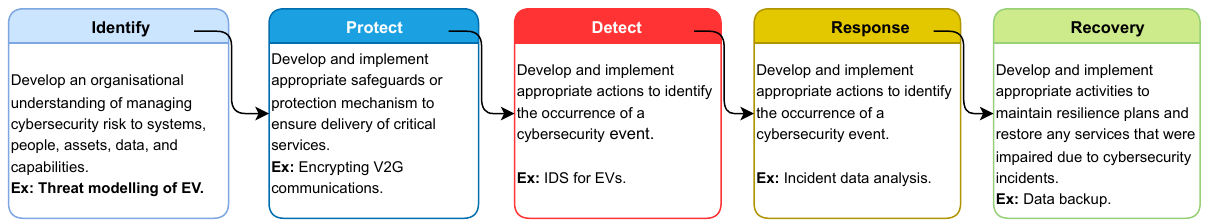}
\caption{Cybersecurity risk management lifecycle core functions.}
\label{fig3b}
\end{figure*} 

\subsection{Cybersecurity Risk Management Lifecycle (CRML) Functions}

Cybersecurity frameworks, such as the NIST Cybersecurity Framework~\cite{NIST-24}, provide comprehensive guidance to organisations on effectively preventing, detecting, and responding to cyberattacks. It conceptualises cybersecurity risk management as a continuous lifecycle structured around five core functions: Identify (Id), Protect (P), Detect (D), Respond (R), and Recover (Re). Figure~\ref{fig3b} briefly defines these functions, which are neither strictly sequential nor entirely independent, but provide a holistic and flexible framework that adapts to various organisational contexts. This strategic categorisation facilitates alignment with an organisation's broader risk management and operational goals while ensuring resilience against evolving cyber threats.  

Existing research studies on cybersecurity of V2G (e.g.,~\cite{Paper-2, Paper-3, Paper-4, Paper-5, Paper-6, Paper-7, Paper-8, Paper-10, Paper-12}) can be categorised and analysed according to these functions of the CRML. These studies propose solutions that address one or more CRML functions. For example, some focus on threat modelling or identifying attack vectors, while others encompass multiple aspects, such as threat modelling and anomaly detection. For example, the authors in~\cite{Paper-6} aim to identify effective attack vectors and robust attack strategies for V2G systems, focusing on EVCS. In contrast,~\cite{Paper-7} proposes the detection and localisation of edge-based adversarial oscillatory load attacks and a distributed mitigation technique. This mitigation approach can accelerate power system recovery in the event of an attack, addressing both the ion and recovery functions of the NIST framework.

\subsection{V2G Components}

As discussed in Section~\ref{sec:2} and illustrated in Figure~\ref{fig1}, a V2G system consists of six key components (i.e., EVs, EVCS, SGN, EMTP, TP and U). Existing work could be categorised in terms of the V2G components they support. These studies propose cybersecurity solutions for one or more of these components. For example,~\cite{Paper-10} focus on securing the deployment of EVCS, while~\cite{Paper-7} presents an edge-based cybersecurity solution for EV and EVCS. Furthermore,~\cite{Paper-17} introduces an efficient mutual authentication scheme for EMTP to improve the security of EV, EVCS, and the SGN.

Some studies selected in this review do not explicitly reference V2G systems but remain highly relevant to this field. For example, a blockchain-based secure demand response management scheme~\cite{Paper-37} has been proposed for the SGN, although it does not explicitly address V2G systems. These studies will be categorised with others (e.g., Others for smart grid-related work). Hence, we will analyse the existing studies using the following categories.

\begin{itemize}
    \item One of the V2G components (e.g., EV or EVCS) or Others (e.g., SGN) only: If the study proposes solution or solutions for one of the V2G components or one of the other systems (e.g., SGN) or components (e.g., smart meter (SM)),
    \item A combination of V2G components (e.g., EV and EVCS) or other components or systems (e.g., EV and SM)
    \item V2G: If the proposed solution or solutions address the cybersecurity of all six components of a V2G system.  
  \end{itemize}
  
This categorisation will help us to understand the components-wise research gaps in V2G cybersecurity and identify future research.

\begin{figure*}[hbt!]
\centering
\includegraphics[width=17.5cm]{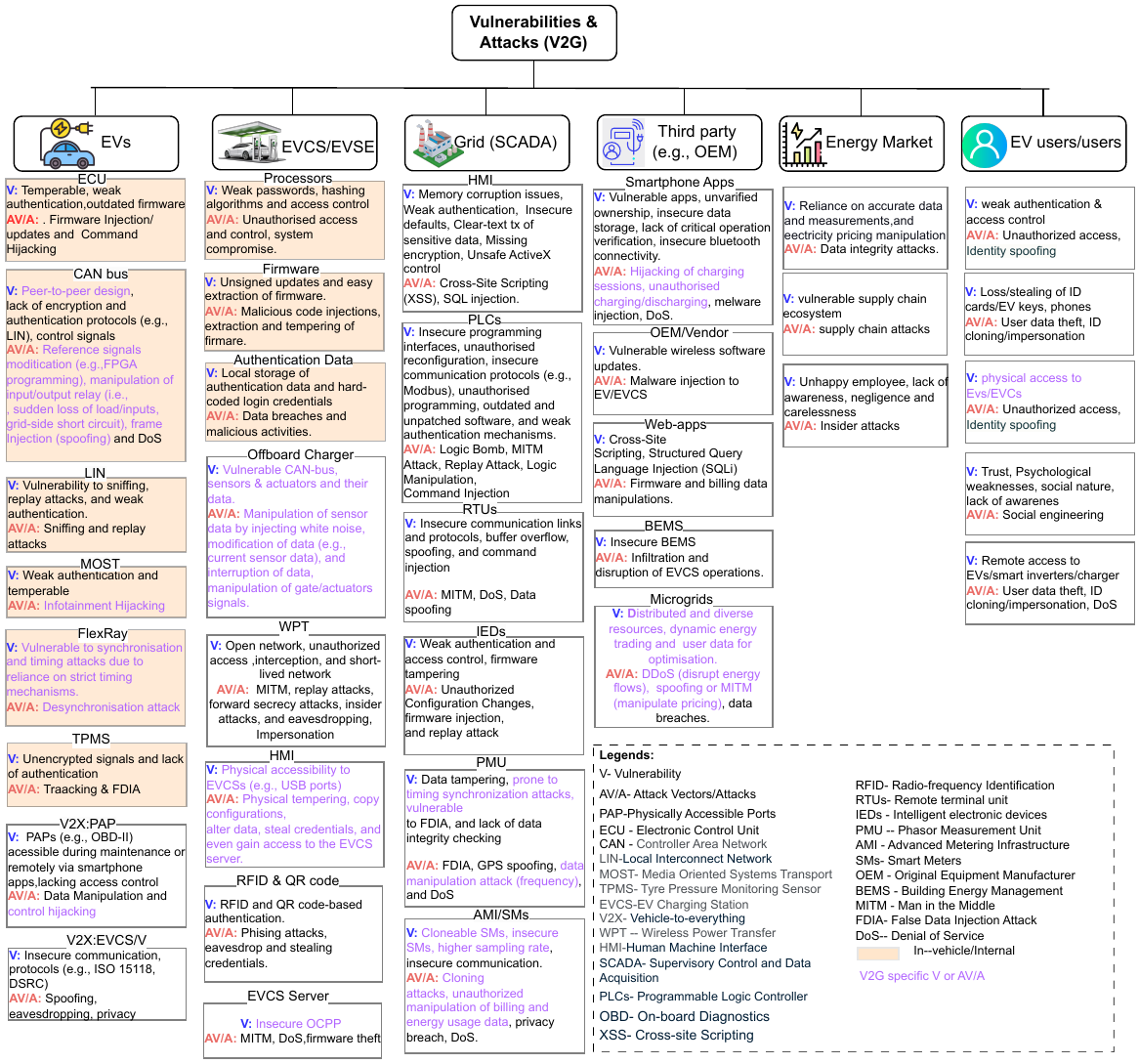}
\caption{Vulnerabilities and their respective attacks in V2G (components-wise and generic vs. unique to V2G).}
\label{fig3c}
\end{figure*}

\subsection{Vulnerabilities and Attack Vectors}

A comprehensive list of vulnerabilities and attack vectors or attacks in V2G systems is essential for a robust, end-to-end (e.g., EV, EVCS, SGN, U) and holistic security strategy that addresses all potential weak points or vulnerabilities in the V2G ecosystem. Researchers and practitioners can better understand the security landscape, prioritise critical risks, and design targeted defences by systematically identifying vulnerabilities and potential attack vectors~\cite{aljohani2024comprehensive, acharya2020cybersecurity,achaal2024study}. 

Identifying unique V2G vulnerabilities is crucial due to their specialised operations, such as bidirectional energy flow and decentralised protocols. Understanding these unique and specific vulnerabilities of V2G and attack vectors or attacks ensures tailor-made cybersecurity measures, prevents critical gaps, improves the reliability of the grid and supports regulatory compliance for safer, interoperable systems~\cite{OBC-attacks2020,ronanki2023electric, mobile-apps-attacks2023}.

Figure~\ref{fig3c} presents a comprehensive list of component-wise vulnerabilities (V) and their respective attacks on V2G, including generic (black text) and unique vulnerabilities (purple text) and attacks on V2G systems. In addition, EV and EVCS-related vulnerabilities and attack vectors or attacks are highlighted as in-vehicle or internal (yellow shaded text) and external vulnerabilities or attacks. For example, an attacker can exploit weak authentication of an ECU (a common and internal vulnerability of EVs, since ECUs are installed inside an EV) to inject firmware (a common and internal attack to an EV) into ECUs. In contrast, FlexRay is a high-speed, deterministic, and fault-tolerant automotive communication protocol that is vulnerable to lack of synchronisation and timing attacks (unique and internal to EVs) due to the reliance on strict timing mechanisms~\cite{OBC-attacks2020,ronanki2023electric, mobile-apps-attacks2023}.

\begin{figure*}[hbt!]
\centering
\includegraphics[width=17.5cm]{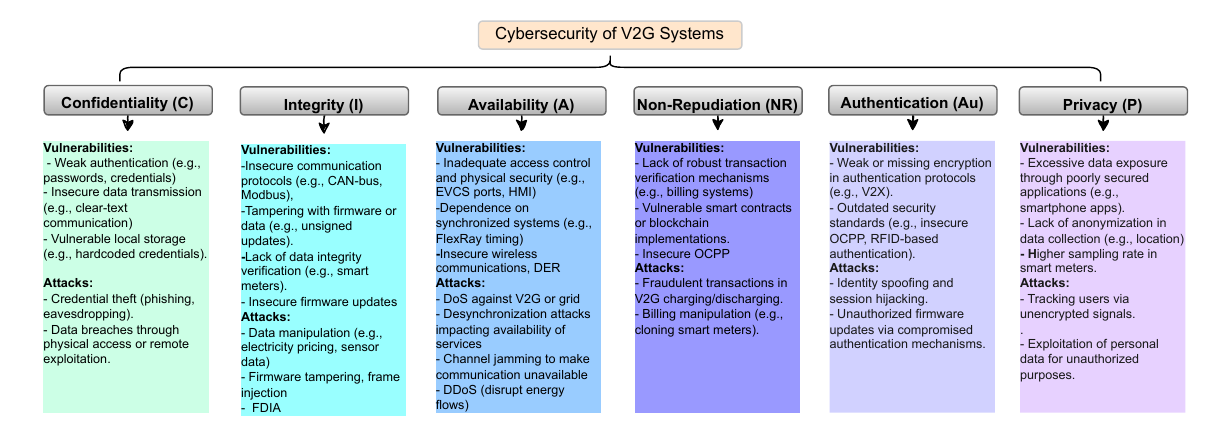}
\caption{Vulnerabilities and their respective attacks categorisation using CIA and others.}
\label{fig3d}
\end{figure*} 

\subsection{Cybersecurity Principles}

Identifying and understanding the core cybersecurity principles addressed in existing studies, including confidentiality (C), integrity (I), availability (A), nonrepudiation (NR), authentication (Au) and privacy (P), are crucial for advancing V2G system research. These principles offer a structured framework for assessing security and identifying gaps in existing solutions. Researchers can evaluate the effectiveness of methodologies by explicitly linking cybersecurity solutions to these principles and pinpointing areas that need further investigation. This approach drives innovation in critical areas, such as quantum-resistant protocols and resilient architectures against DDoS attacks. In addition, focusing on these principles facilitates the design of comprehensive solutions that balance security goals with scalability, robustness, and AI explainability. Such alignment ensures that future research addresses immediate threats and contributes to the long-term security and trustworthiness of V2G systems in an evolving threat landscape~\cite{chandwani2020cybersecurity, aljohani2024comprehensive}.

Figure~\ref{fig3d} outlines a list of vulnerabilities and their associated attacks on V2G systems, focusing on six core cybersecurity principles: confidentiality (C), integrity (I), availability (A), non-repudiation (NR), authentication (Au) and privacy (Pr). For example, an attacker could compromise the confidentiality of an EV user's credentials for the EVCS mobile app by exploiting vulnerabilities such as weak authentication mechanisms or poor password security~\cite{mobile-apps-attacks2023}. Similarly, the integrity of sensor data on an off-board charger (OBC) could be compromised by manipulating sensor data through an insecure CAN bus~\cite{OBC-attacks2020}.

\subsection{Methodology: Key Technologies}

Identifying the key technologies used in V2G cybersecurity solutions is important to understand their capabilities, effectiveness, and limitations. This understanding informs future research, encourages innovation, and ensures that solutions align with evolving threats and requirements. In the following, we present a brief description of four key technologies (i.e., Blockchain, AI, Encryption, and Hardware-based). 

\begin{itemize}
    \item Blockchain (BC): Blockchain is the most widely used technology in V2G cybersecurity~\cite{miglani2020blockchain, Paper-14}, which provides decentralised, immutable ledger systems for secure V2G transactions, enabling trust between multiple parties without intermediaries. It features smart contracts for automated security policy enforcement and consensus mechanisms for distributed validation, which are particularly valuable in charging and billing processes. 
    
    \item Artificial intelligence (AI): AI enables data-driven real-time anomaly detection, predictive threat analysis, and adaptive security responses in V2G networks. AI (especially ML and DL), can identify complex attack patterns, optimise security configurations, and provide automated incident response capabilities on distributed V2G infrastructure~\cite{Paper-5, Paper-7}.

    \item Encryption: Cryptographic protocols secure data confidentiality and integrity in V2G communications~\cite{Paper-3, Paper-13}. These include symmetric/asymmetric encryption, digital signatures, and key management systems, which protect sensitive information during V2G interactions and financial transactions.
    
    \item Hardware-based: Hardware-based solution techniques, including Physical Unclonable Functions (PUF) and trusted platform modules, provide hardware-level security guarantees for device authentication and data protection~\cite{Paper-12, Paper-15}. These solutions offer tamper resistance and unique device fingerprinting capabilities essential for secure V2G endpoint protection.
    
    \item Miscellaneous Methodologies (MISC): Many existing studies (e.g.,~\cite{Paper-3, Paper-18, Paper-19, Paper-36, Paper-39}), encompassing a wide range of approaches, including simulation, optimisation, and control theory. For example, ~\cite{Paper-39} used simulation to evaluate the resilience of V2G systems under various attack conditions and~\cite{Paper-3, Paper-21, Paper-33, Paper-134, Paper-178} studied mechanisms based on control theory to improve cybersecurity features in V2G networks. 
\end{itemize}

\subsection{Performance or Quality Metrics}

This subsection briefly presents a list of quality or performance metrics/features of the existing proposed V2G cybersecurity solutions. An understanding and inclusion of these metrics/features is crucial because:

\begin{itemize}
    \item V2G systems are critical infrastructures that require robust, scalable, and efficient solutions to handle the growing adoption of EVs.
    
    \item These features directly impact the real-world deployability and long-term viability of security solutions.
    
    \item Future-proofing against quantum computing threats and AI-related vulnerabilities is essential for long-term grid security.
    
    \item Features such as efficiency and lightweight design are essential for real-time operations in resource-constrained environments (e.g., wireless IoT devices).
    
    \item Grid stability and reliability of the grid need resilient and reliable security mechanisms.
\end{itemize}

A list of key performance or quality metrics of existing V2G cybersecurity solutions is breifly defined below.

\begin{enumerate}
    \item \textbf{AI Assurance (AIA)} ensures that AI systems in V2G cybersecurity are explainable, safe, reliable, and aligned with operational and ethical goals.
    \item \textbf{Communication Efficiency (CoE) }optimises data exchange in V2G systems, minimising latency and bandwidth usage while ensuring security.
    \item \textbf{Computational Efficiency (CE)} minimises resource consumption and processing time for V2G cybersecurity solutions, improving performance.
    \item \textbf{Energy Efficiency (EE)} reduces power consumption in V2G operations, balancing cybersecurity and system sustainability.
    \item \textbf{Lightweight (LW) }ensures solutions require minimal memory or computational resources, which is suitable for devices with limited resources.
    \item \textbf{Quantum-safe (QS)} solutions protect V2G systems against quantum computing threats through robust cryptographic mechanisms.
    \item \textbf{Resilience (Re)} is the maintenance of essential security functions despite failures or attacks.
    \item \textbf{Robustness (R)} tolerates adverse conditions, such as hardware faults or unexpected inputs, without compromising security.
    \item \textbf{Scalability (Sc)} ensures that solutions are seamlessly adapted to increase the size or complexity of the system in V2G deployments.
    \item \textbf{Trustworthiness (T)} builds user confidence in V2G systems by ensuring transparency, security, and reliability.
    \item \textbf{Traceability (Tr)} enables the tracking of data and actions within V2G systems, improving accountability and forensic analysis.
\end{enumerate}

\begin{figure*}[hbt!]
\centering
\includegraphics[width=15cm]{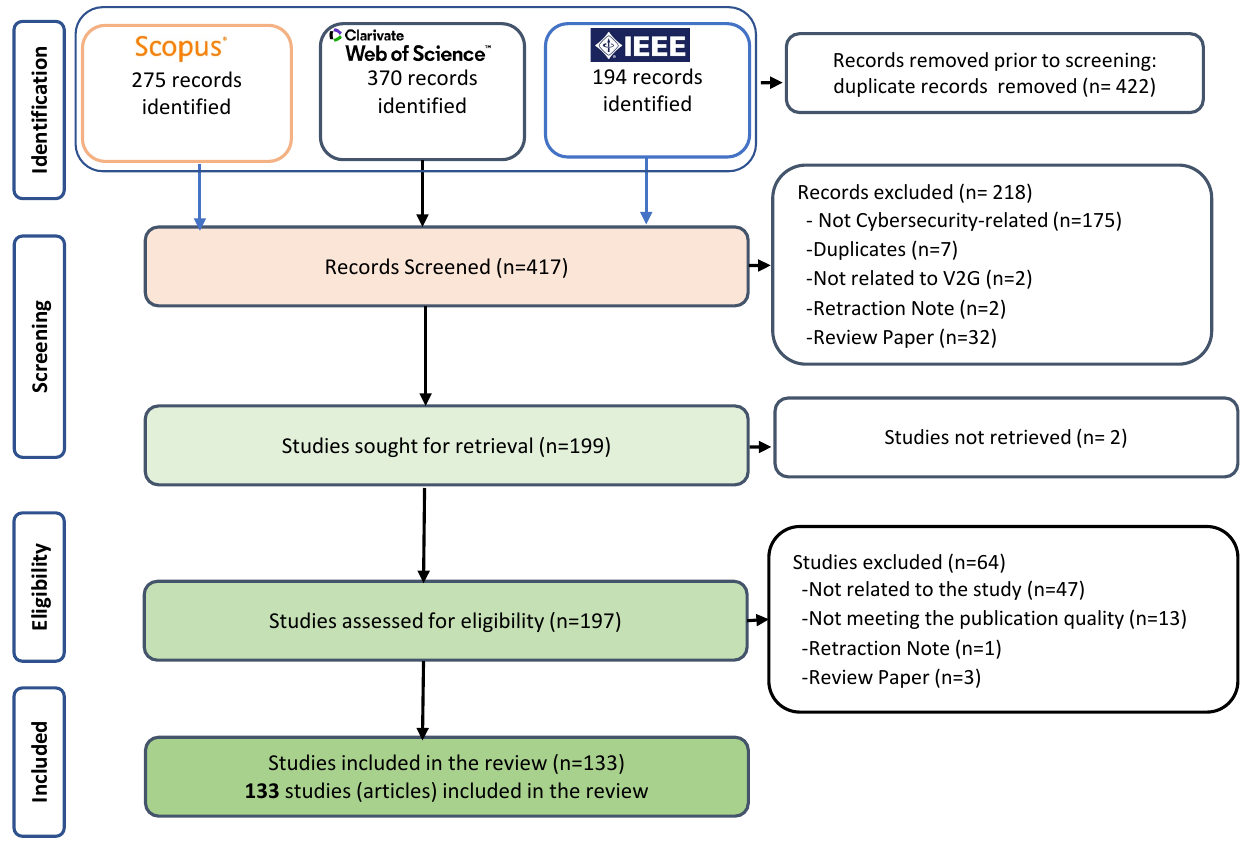}
\caption{PRISMA-based existing work selection strategies.}
\label{fig4a}
\end{figure*}

\section{Existing Work Selection Strategies}
\label{sec:3b}

We used the recent PRISMA guideline~\cite{PRISMA-2020} to systematically filter our search results. We identified 133 qualified scientific articles for the review study (Figure~\ref{fig4a}). The selection started with 275, 370 and 194 articles from the Scopus, Web of Science, and IEEE databases search results, respectively, using the following keywords:

\begin{itemize}
    \item "Electric Vehicle" or EV or " Electric Vehicles (EVs)" or "V2G Networks " or " Vehicle-To-Grid, V2G Networks " and
    \item "Cyber Security" or " Cybersecurity" or " Security " or " Cyber Attacks" or " Fault Data Injection" or "False Data Injection Attacks" (Topic) and
    \item "Power Grid" or  " Smart Grid", or " Charging Infrastructure", or "EV Charging Station"
\end{itemize}

Our study included all publications from 2019 to June 19, 2024. 422 duplicate entries were removed from the two databases (Figure~\ref{fig4a}). Next, we screened the titles and abstracts of the publications, removing conference papers, editorials, perspectives, technical documents, other review and survey studies, and unrelated studies that did not pertain to V2G cybersecurity. This process resulted in the identification of 197 scientific studies or potential inclusion in this review study. Finally, after reviewing the list following the exclusion criteria (as illustrated in Figure~\ref{fig4a}, including excluded articles that do not meet the quality of the publication, such as unknown publisher or predatory journals or publishers~\cite{IsMDPIap61:online}), we ultimately found 133 journal articles to include in this study.

\section{Result and Discussion}
\label{sec:4}

\begin{figure}[hbt!]
\centering
\includegraphics[width=8cm]{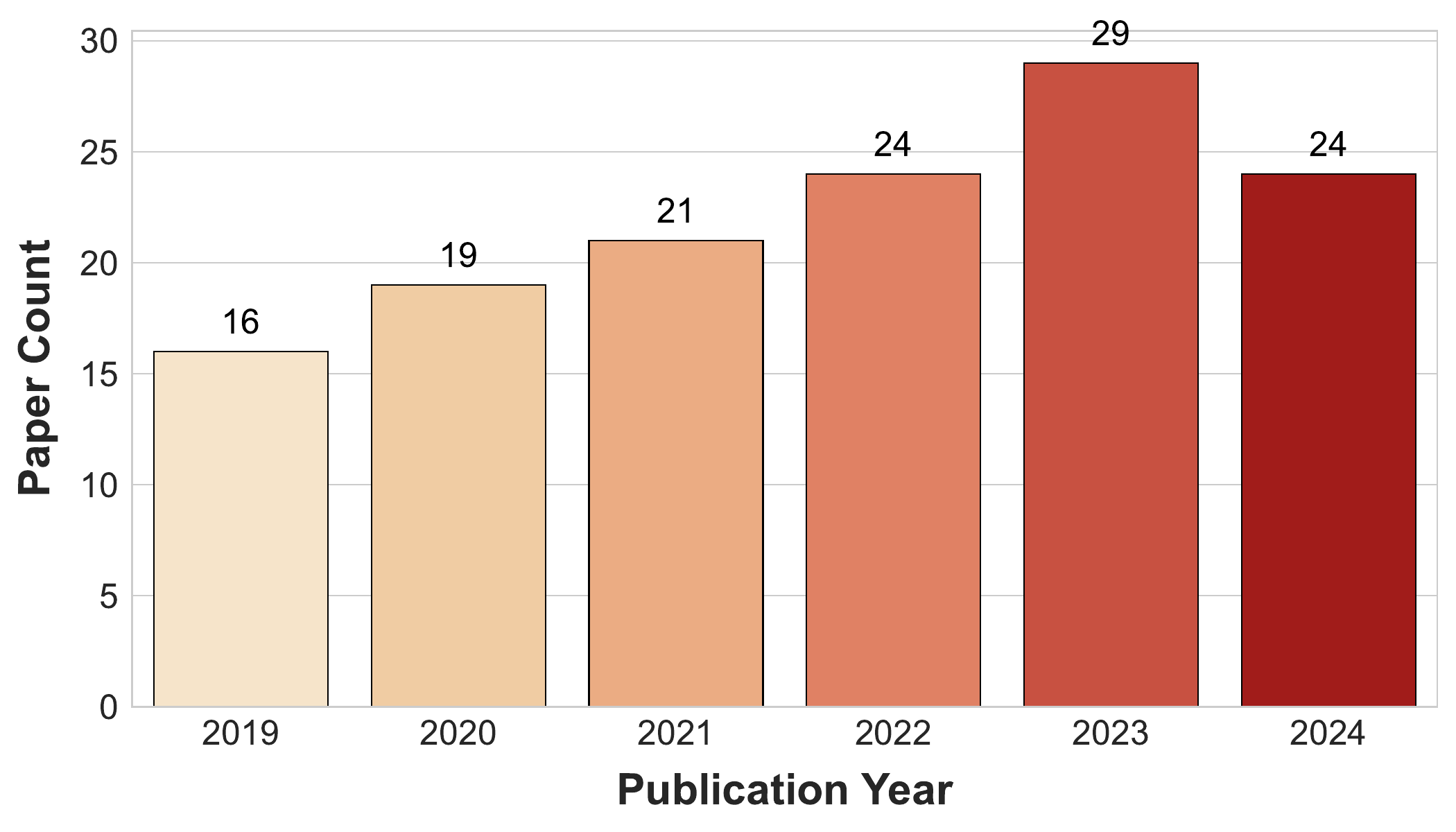}
\caption{Yearly Publications on V2G Cybersecurity.}
\label{fig5a}
\end{figure}

Cybersecurity in V2G systems is a critical and rapidly evolving area of research (Figure~\ref{fig5a}), especially with the increasing adoption of EV and the surge of cyberattacks on critical infrastructure. Recent incidents, such as the Ukraine Power Grid Attack (2015, 2016)~\cite{CyberAtt19:online} and the Colonial Pipeline Ransomware Attack (2021)~\cite{Cybersec19:online}), highlight these concerns. Numerous studies~\cite{Paper-2, Paper-3, Paper-4, Paper-5, Paper-6, Paper-7, Paper-8, Paper-10, Paper-12} have been published on V2G cybersecurity, and as shown in Figure~\ref{fig5a}, the number of publications is gradually increasing each year. For example, in less than six months of 2024, 24 articles were published compared to 29 in 2023. These studies vary in their focus on CRML functions, V2G components, vulnerabilities, and the attacks addressed, along with the key technologies used in their proposed solutions. In the following subsections, we evaluate existing V2G cybersecurity studies based on the features described in Section~\ref{sec3}, specifically, how critically they addressed these key features.

\subsection{V2G Components}
\label{5.2}

\begin{figure}[hbt!]
\centering
\includegraphics[width=8cm]{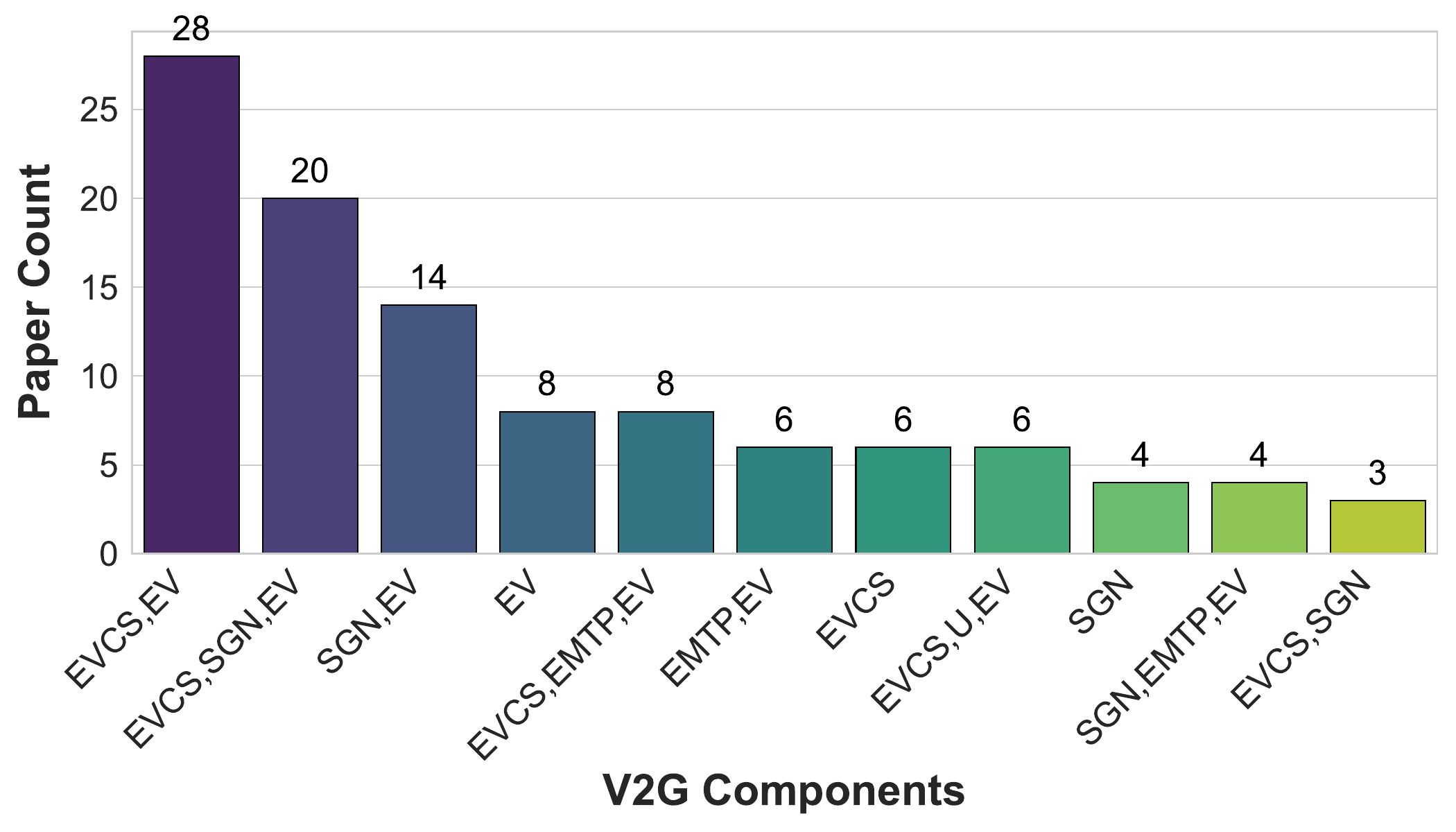}
\caption{Distribution of papers addressing specific combinations of V2G components.}
\label{fig5d}
\end{figure}

\begin{figure}[hbt!]
\centering
\includegraphics[width=8cm]{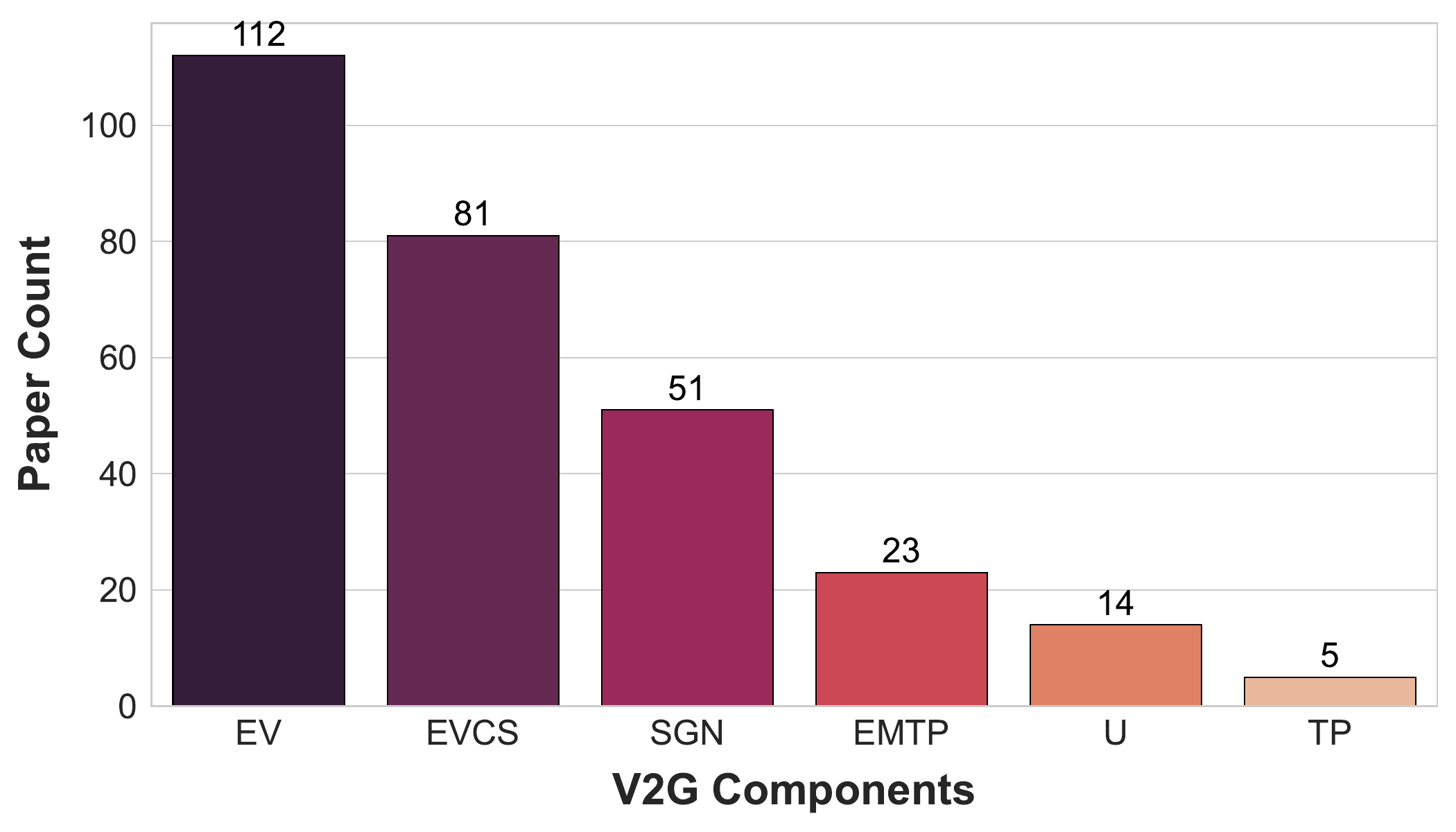}
\caption{Distribution of papers addressing individual V2G component.}
\label{fig5e}
\end{figure}

As explained earlier, A V2G cyber-physical system (CPS) includes six key components: EV, EVCS, SGN EMTP, TP, and U, each of which is complex. Existing studies often focus on specific components/subcomponents or a combination of components. Figure~\ref{fig5d} illustrates the distribution of paper counts with respect to specific combinations of V2G components, while Figure~\ref{fig5e} highlights the studies that address each component. As shown, various combinations of V2G components (V2G~\cite{Paper-15, Paper-39, Paper-42}, EMTP, EVCS, EV and SGN~\cite{Paper-46, Paper-47}, EVCS, EMTP and EV~\cite{Paper-19, Paper-87, Paper-113}, SGN and EV~\cite{Paper-27, Paper-33, Paper-68}, and EV~\cite{Paper-36} or SGN~\cite{Paper-2}) have been explored, the most common being EVCS and EV (28 out of 133 studies), followed by EVCS, EV, and SGN (20 studies). In particular, some combinations, such as EVCS (based on wireless power transfer) and EV~\cite{Paper-14}, were only considered in single studies. Furthermore, eight studies (e.g.~\cite{Paper-13, Paper-36, Paper-145}) focused solely on the cybersecurity of EVs, while four investigated SGN-related cybersecurity~\cite{Paper-2, Paper-37, Paper-66, Paper-178}. However, none addressed end users or EV users (U) from a behavioural perspective, despite humans being the weakest link in security solutions, accounting for more than 80\% of data breaches~\cite{ovelgonne2017understanding}. Lastly, eight studies~\cite{Paper-15, Paper-39, Paper-42, Paper-44, Paper-65, Paper-98, Paper-119, Paper-126} that claim to consider system-wide V2G inadvertently overlooked one or two of the six identified components, with~\cite{Paper-42} proposing a twin-enabled digital security framework and~\cite{Paper-15} introducing a PUF-based authentication scheme without addressing EMTP.


As illustrated in Figure~\ref{fig5e}, EV has received the most significant research attention, with 112 articles proposing solutions for the cybersecurity of EV. For example,~\cite{Paper-13} proposed a solution only for the cybersecurity of EV, and many other studies considered EV alongside other V2G components (for example, the solution of EVCS, EMTP and EV~\cite{Paper-19}, SGN and EV~\cite{Paper-27}). EVCS follows EV, and 81 studies proposed security solutions for EVCS in the context of V2G systems.  

Moderate attention has been paid to SGN and Energy Market and Trading Platforms (EMTP), with 51 and 23 articles, respectively. However, U and third-party (TP) entities have received limited focus, with just 14 and 5 articles. In particular, some studies examined malicious activities by EV users~\cite{Paper-167, Paper-195} and interactions with TP~\cite{Paper-8, Paper-34, Paper-43, Paper-47, Paper-112}. Finally, several studies(e.g.,~\cite {Paper-87, Paper-117, Paper-147, Paper-155}) emphasise the cybersecurity of aggregators, including local aggregators, indicating a growing need for comprehensive security strategies across this evolving landscape. 

\begin{figure}[hbt!]
\centering
\includegraphics[width=8cm]{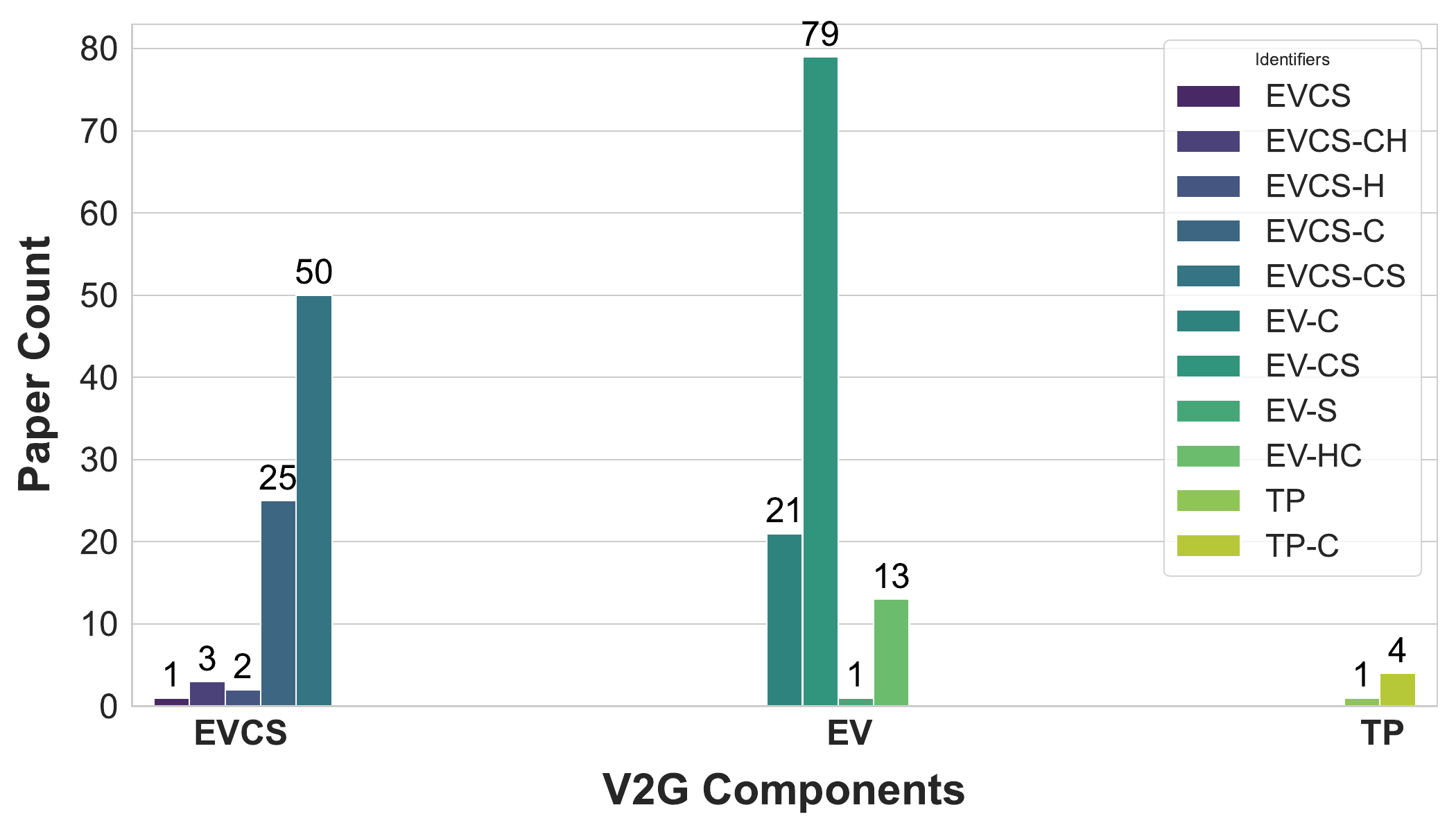}
\caption{Distribution of papers addressing V2G sub-components.}
\label{fig5f}
\end{figure}

Figure~\ref{fig5f} illustrates the distribution of published articles across V2G subcomponents, particularly EVCS, EV and TP. Generally, subcomponents such as communication (EV-C or EVCS-C), including communication protocols (e.g., OCCP, OpenADR), software (EV-S or EVCS-S), including mobile apps, hardware (EV-H or EVCS-H), communication and software (EV-CS or EVCS-CS), and communication and hardware (EV-CH or EVCS-CH) of EV and EVCS are the target of attackers. In the case of EVs, most studies (79/112) (e.g.~\cite{Paper-5, Paper-12, Paper-18}) have focused on the cybersecurity of communication components (EV-C), followed by C and S (EV-CS), with 21 articles (e.g.,~\cite {Paper-87, Paper-116, Paper-117}) and 13 articles (e.g.,~\cite{Paper-36, Paper-120, Paper-133}) focused on the C and H (EV-CH) of EVs. In contrast, most studies related to EVCS (50/81) (e.g.,~\cite{Paper-5, Paper-12, Paper-18}) have focused on the C and S aspects (EVCS-CS) of EVCS, followed by the C aspects (EVCS-C) only with 25 studies.

\begin{figure}[hbt!]
\centering
\includegraphics[width=8cm]{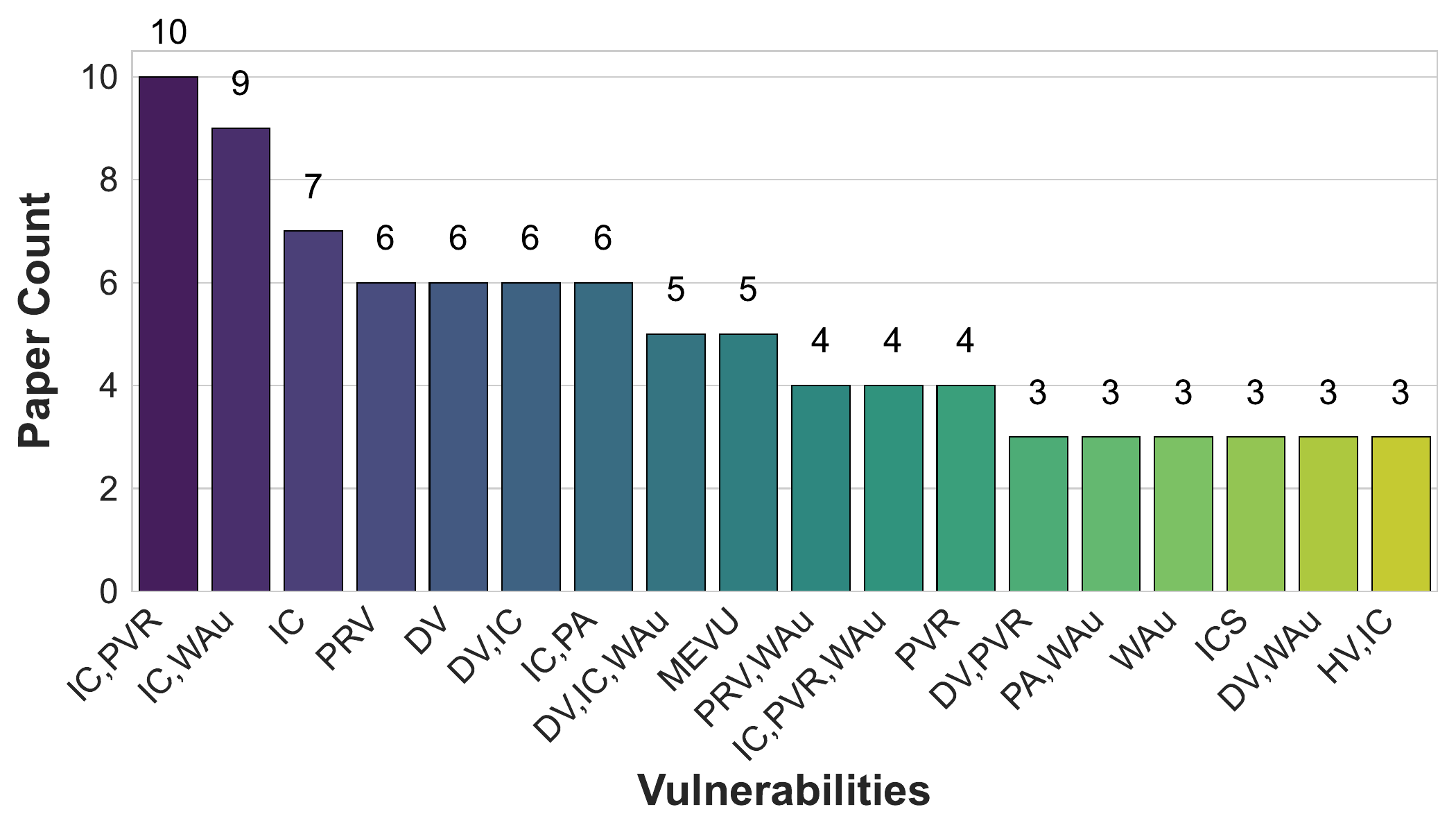}
\caption{Distribution of papers addressing specific combinations of vulnerabilities.}
\label{fig5g}
\end{figure}

\begin{figure}[hbt!]
\centering
\includegraphics[width=8cm]{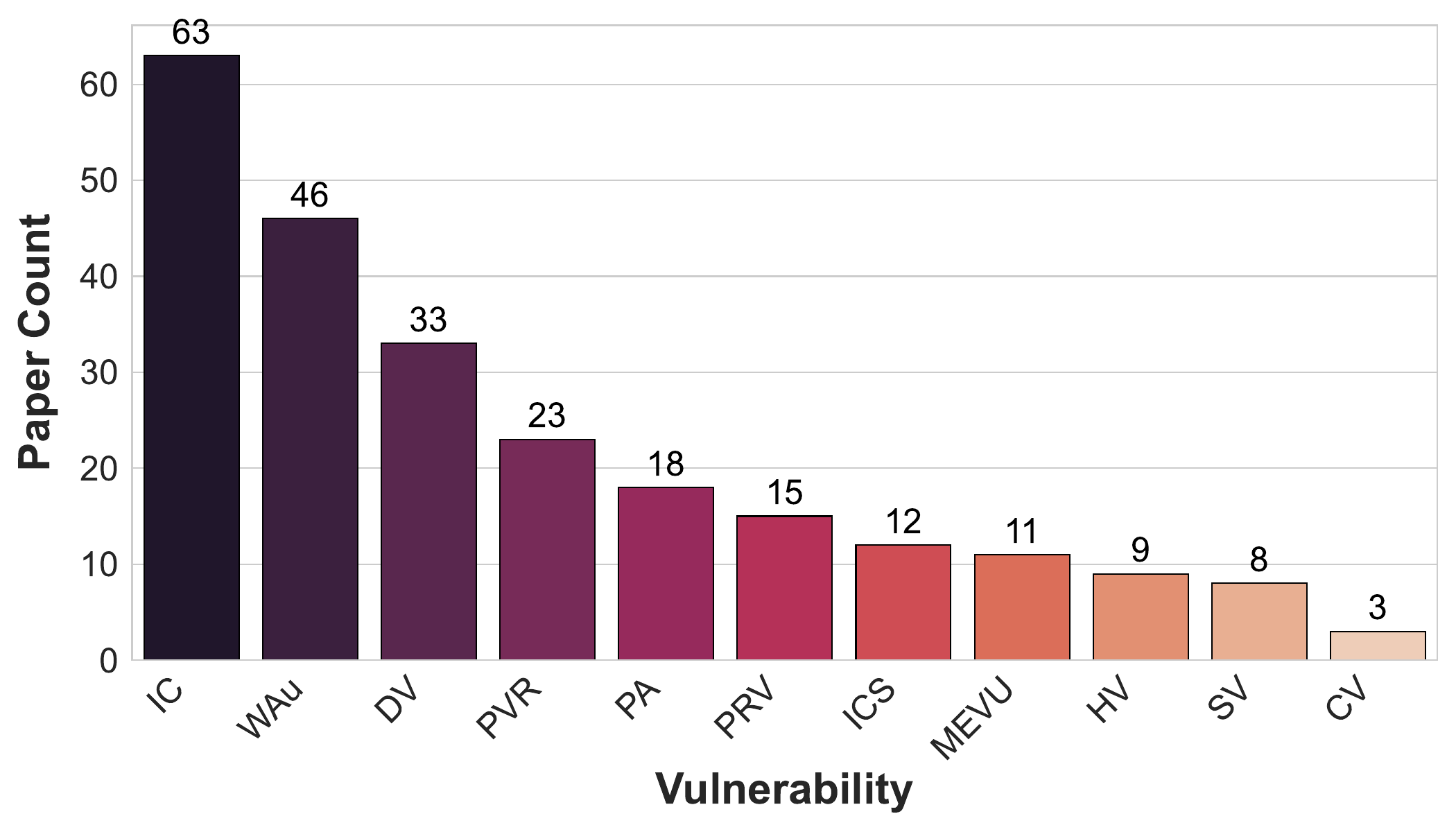}
\caption{Distribution of papers addressing individual Vulnerability.}
\label{fig5h}
\end{figure}

\begin{figure}[hbt!]
\centering
\includegraphics[width=8cm]{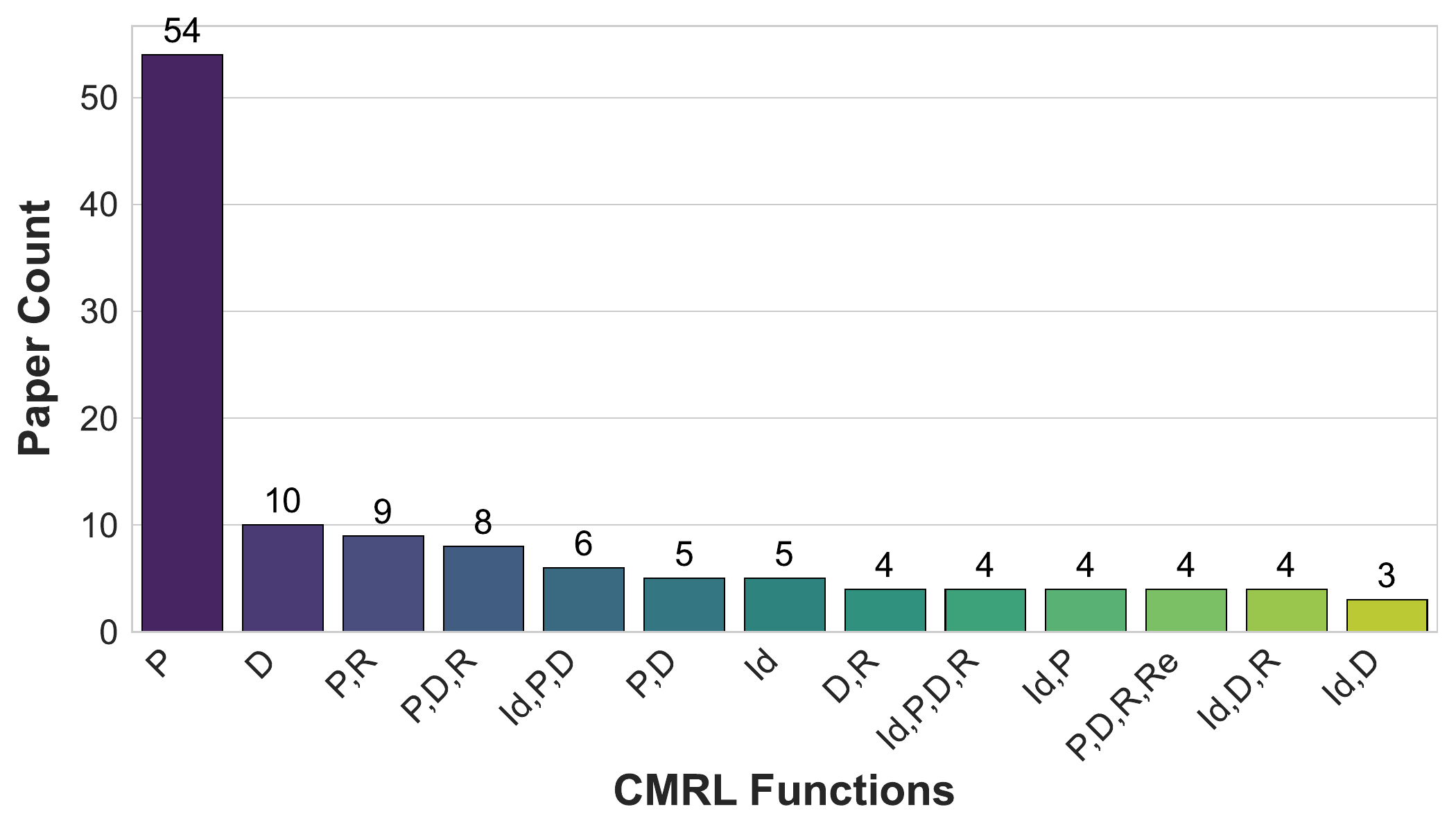}
\caption{Distribution of papers addressing specific combinations of CRML functions.}
\label{fig5b}
\end{figure}

\begin{figure}[hbt!]
\centering
\includegraphics[width=8cm]{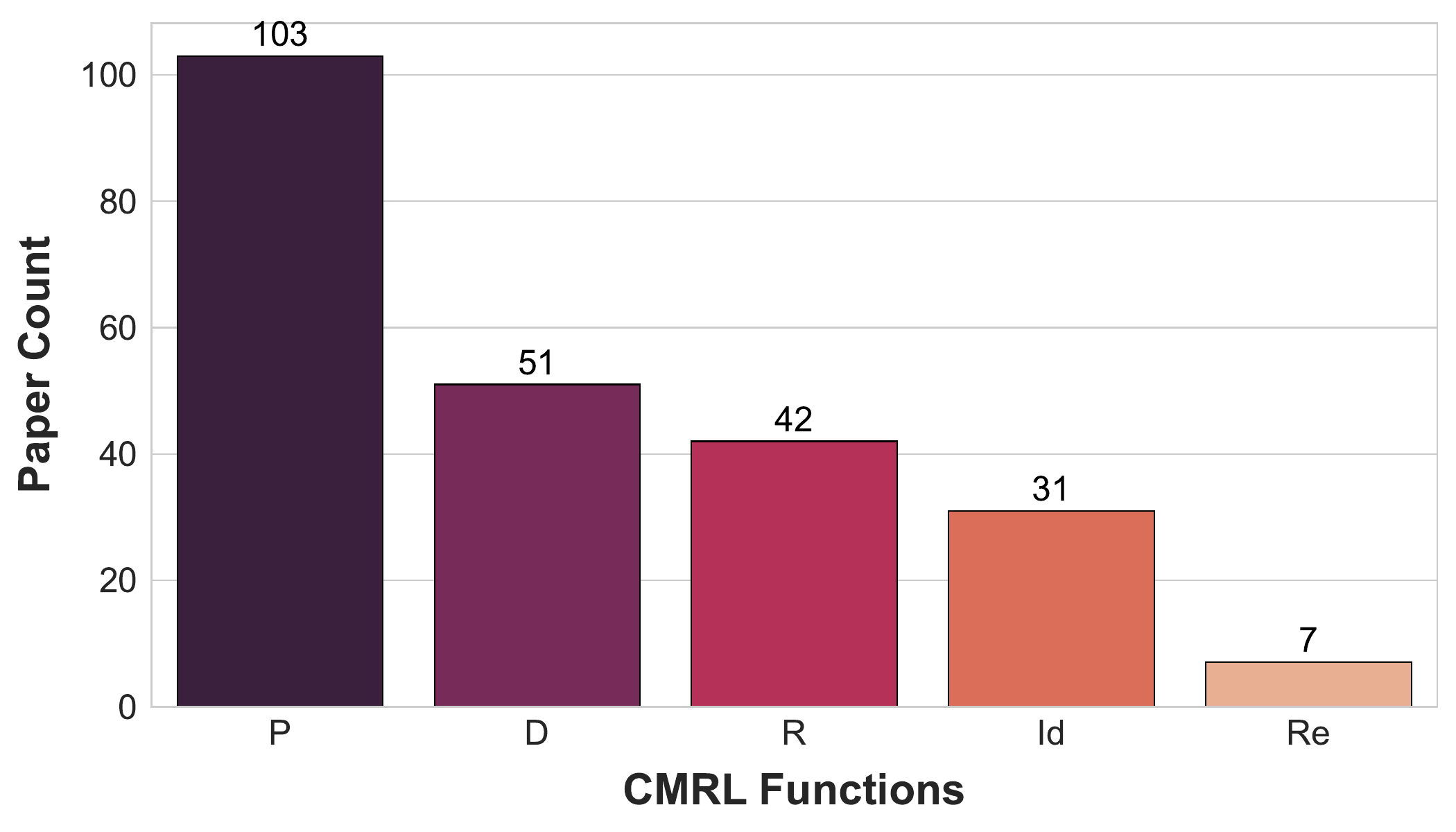}
\caption{Distribution of papers addressing individual CRML functions.}
\label{fig5c}
\end{figure}

\subsection{CRML Functions}

Figures~\ref{fig5b} and~\ref{fig5c} provide an overview of the distribution of published articles in the five key CRML functions. These functions are critical to understanding the cybersecurity landscape of V2G systems, which are increasingly integral to modern smart grid infrastructures.

Figure~\ref{fig5b} illustrates the distribution of paper counts between the CRML functions, while Figure~\ref{fig5c} highlights the frequency with which each function has been considered. Protection (P) has received the most significant research attention, with 103 articles focusing on protecting V2G systems against cyber threats. Of these, 54 studies (53\%) exclusively address protection, while the remaining 49 combine protection with other functions, such as detection (D) and response (R). For example, studies~\cite{Paper-2, Paper-37} propose solutions solely to protect the grid from cyberattacks, while others, including~\cite{Paper-4, Paper-12, Paper-22}, focus on the security of EVs, EVCSs, and related components.

Detection (D) follows P, and 51 studies explore strategies to identify cyber threats. Among these, 10 studies focus solely on detection. Some studies, for example,~\cite{Paper-157, Paper-158} address P and D, and~\cite{Paper-7, Paper-30} integrate D with R, highlighting a multilayered approach to cybersecurity.

Response (R) and identification (Id) have received moderate research attention, with 42 and 31 articles, respectively. Only two studies focus solely on R, while five exclusively address I. For example,~\cite{Paper-74} focuses on responding to the grid's load-altering attacks. Other studies, such as~\cite{Paper-102, Paper-135, Paper-138}, explore cyber attack responses in combination with other CRML functions. Identifying potential threats in V2G systems, whether through threat modelling (TM) or other approaches, is crucial to adequately address protection, detection, and response. However, no study has conducted a comprehensive system-wide threat modelling analysis for V2G systems. Some works, such as~\cite{Paper-6, Paper-10, Paper-199}, analyse threats in EVCS, while~\cite{Paper-152, Paper-173} extend their focus to EVs, EVCS, and the grid. In addition,~\cite{Paper-64} identified the attack surface of EV charging mobile applications. 

Recovery (Re) has received the least attention, with only seven studies~\cite{Paper-109, Paper-146, Paper-149, Paper-162, Paper-164, Paper-195} addressing it alongside other functions, none focus exclusively on it. This indicates a significant gap in research on the restoration of V2G systems after a cyberattack.

\subsection{Vulnerabilities}

Most of the vulnerabilities and related attacks discussed in existing studies are summarised in Figure~\ref{fig3c}. Figure~\ref{fig5g} shows the distribution of paper counts based on the specific combination of vulnerabilities they address, while Figure~\ref{fig5h} highlights studies that focus on each vulnerability. Most studies tackle more than one vulnerability, with various combinations, such as privacy-related vulnerabilities (PRV)~\cite{Paper-66, Paper-68, Paper-71}, insecure communication (IC) and PRV~\cite{Paper-127, Paper-129}, and data-related vulnerabilities (DV) and IC~\cite{Paper-107, Paper-162}. The most frequent combinations include PVR (10 out of 133 studies) and PRV with IC, followed by IC with weak authentication (WAu) and IC with DV, with 9, 7, and 6 studies, respectively. In addition, DV~\cite{Paper-45, Paper-72}, IC with physical access (PA)~\cite{Paper-3, Paper-16}, and PRV with WAu~\cite{Paper-87, Paper-99} were each addressed in 6 studies. However, 32 of the 49 vulnerabilities, such as software-related vulnerabilities (SV)~\cite{Paper-14, Paper-64} and hardware-related vulnerabilities (HV)~\cite{Paper-199}, along with some combinations such as DV with PA~\cite{Paper-124} and control-related vulnerabilities (CV) with HV~\cite{Paper-109, Paper-146}, were only covered in one or two studies and are not shown in the figure.

As shown in Figure~\ref{fig5h}, most studies (63 out of 133) focused on vulnerabilities related to IC (e.g.,~\cite{Paper-29, Paper-33, Paper-160}), highlighting the need for secure communication channels. Furthermore, 12 studies, including~\cite{Paper-7, Paper-36, Paper-99}, examined both IC and IC protocols (ICS), specifically pointing out weaknesses in V2G-related communication protocols such as OCCP and OpenADR. Following IC, vulnerabilities related to WAu, PRV, and DV were addressed in 46 (e.g.,~\cite{Paper-13, Paper-34, Paper-98}), 38 (e.g.,~\cite{Paper-66, Paper-68}), and 33 studies (e.g.,~\cite{Paper-45, Paper-70}), respectively.

Physical access (PA) to EV or EVCS (e.g.,~\cite{Paper-7, Paper-8} control-related vulnerabilities (e.g., instability of load frequency control (LFC)~\cite{Paper-109}) are unique to cyber-physical systems, including V2G systems. These vulnerabilities were addressed in 18 and 3 studies, respectively. Although behaviour-related vulnerabilities of EV users remain unexplored, 11 studies have focused on malicious EV users (MEVU). Hardware and software vulnerabilities were discussed in 9 (e.g.,~\cite{Paper-14, Paper-64}) and 8 (e.g.,~\cite{Paper-14, Paper-64}) studies, respectively. Lastly, other vulnerabilities, including uncertainties in power generation, demand response, and vulnerable buses, were mentioned in one or two studies but are not represented in the figure.

\subsection{Attacks}

\begin{figure}[hbt!]
\centering
\includegraphics[width=8cm]{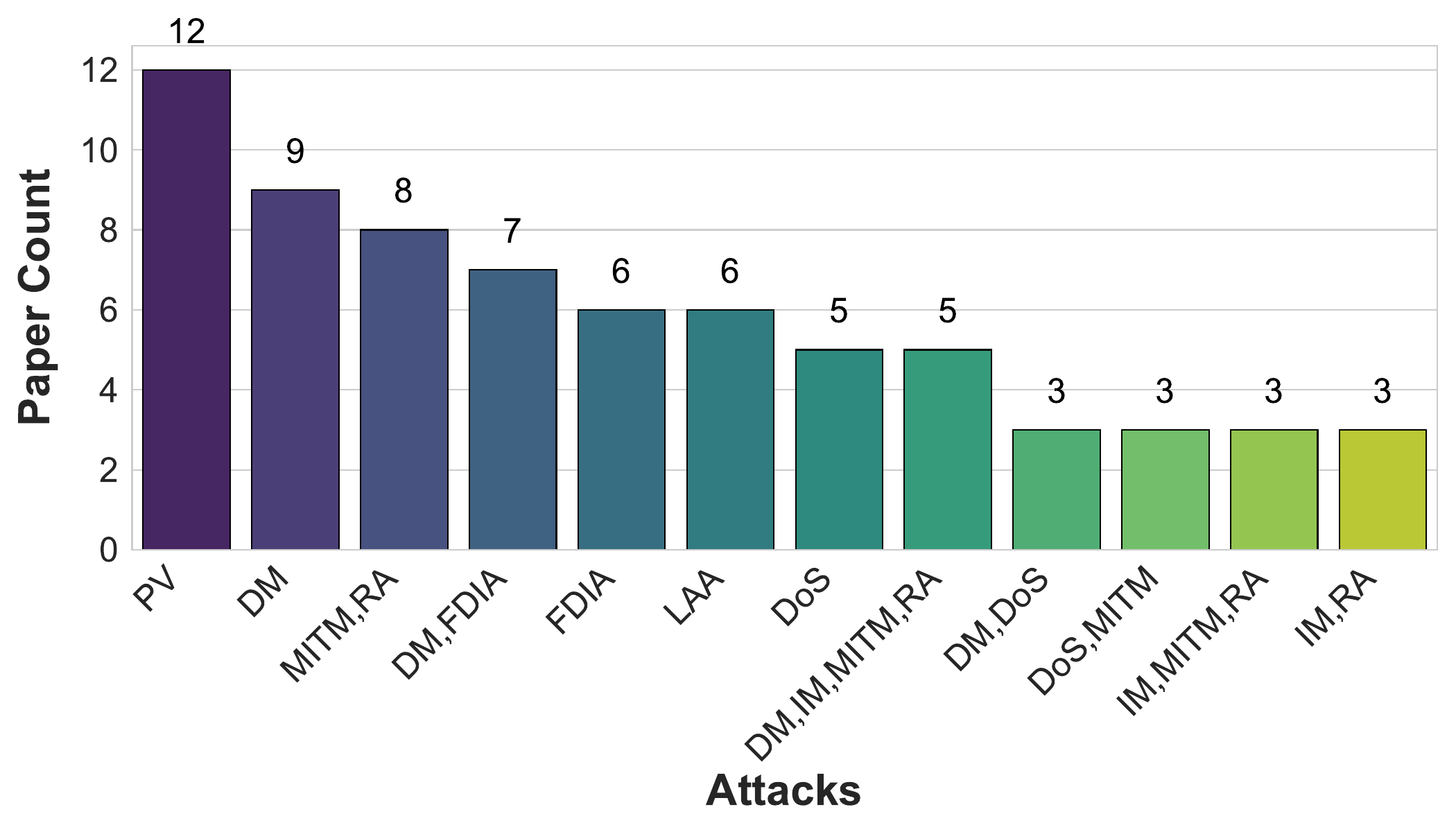}
\caption{Distribution of papers addressing specific combinations of attacks.}
\label{fig5i}
\end{figure}

\begin{figure}[hbt!]
\centering
\includegraphics[width=8cm]{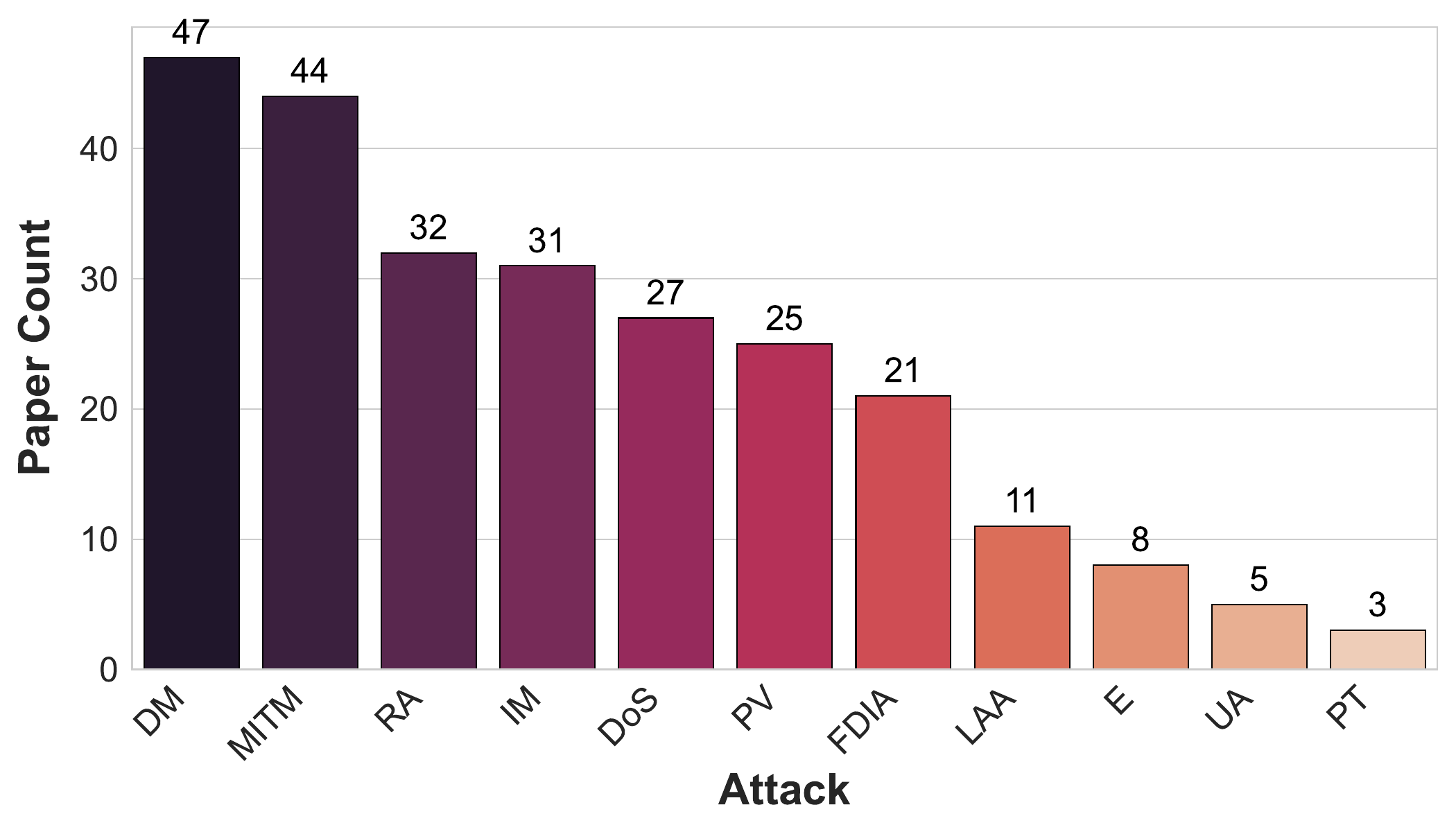}
\caption{Distribution of papers addressing individual attack.}
\label{fig5j}
\end{figure}

Cyber attacks on V2G systems are varied (Figure~\ref{fig3c}), including those specific to V2G (e.g., false data injection) and general attack types (e.g., denial of service). Figure~\ref{fig5i} illustrates the distribution of studies based on the specific combination of attacks that they cover, while Figure~\ref{fig5j} highlights the focus of studies on individual attacks.

Figure~\ref{fig5i} indicates that most studies (86 out of 133) examine multiple attacks, often in various combinations. These include data modification or manipulation (DM) along with DoS~\cite{Paper-97, Paper-151}, DM and eavesdropping (E)~\cite{Paper-145}, and impersonation (Im), MITM, and replay attack (RA)~\cite{Paper-138, Paper-161}. The remaining 47 studies (e.g.,~\cite{Paper-7, Paper-13, Paper-61}) primarily focus on a single attack. The most commonly addressed attacks are privacy violation (PV) (12 out of 133 studies, e.g.,~\cite{Paper-44, Paper-194}) and DM (9 out of 133, e.g.,~\cite{Paper-37, Paper-46}).

In addition, combinations of MITM and RA, DM and FDIA have been addressed in 8 (e.g.,~\cite{Paper-94, Paper-127}) and 7 (e.g.,~\cite{Paper-22, Paper-27}) studies, respectively. Both FDIA and LAA are significant V2G or grid-specific threats, each discussed in 6 studies. For example,~\cite{Paper-61} focuses on FDIA, while~\cite{Paper-7} addresses coordinated oscillatory load attacks (a variant of LAA). DoS and a combination of DM, Im, MITM, and RA were covered by 5 studies each. For example,~\cite{Paper-40}, which proposed an ECC-based authentication framework, and~\cite{Paper-30} developed a DoS attack-resilient grid frequency regulation scheme. Four additional combinations of attacks (DoS and MITM, Im, MITM and RA, DM and DoS, and Im and RA) had 3 studies addressing each. Finally, 53 of the 65 different attacks (e.g., time delay attack (TDA), RA, Malware Injection (MI)) or their combinations (e.g., LAA with phishing (Ph) and DM with Im) were primarily covered in one or two studies. For example,~\cite{Paper-136} suggested a solution to reduce SMS phishing (SMiShing) and LAA, while~\cite{Paper-36} used control theory to mitigate TDA.

Figure~\ref{fig5j} shows that most studies (47 out of 133) and (44 out of 133) focused on DM and MITM, highlighting the need for secure authentication and communication. In particular,~\cite{Paper-37} used blockchain to reduce unauthorised modifications of demand response data, while~\cite{Paper-94} used control theory to address MITM and replay attacks on power grids. Following DM and MITM, RA, Im, DoS, and PV were explored in 32 (e.g.,~\cite{Paper-18, Paper-147}), 31 (e.g.,~\cite{Paper-2, Paper-13}), 27 (e.g.,~\cite{Paper-33, Paper-109}), and 25 (e.g.,~\cite{Paper-66, Paper-174}) studies, respectively.

 V2G or grid-specific attacks, such as FDIA and LAA, have been the focus of 21 (e.g.,~\cite{Paper-107, Paper-146}) and 11 (e.g.,~\cite{Paper-21, Paper-72}) studies, respectively. Research on eavesdropping, unauthorised access (UA) and physical tampering (PT) was limited, with 8 (e.g.,~\cite{Paper-21, Paper-72}), 5 (e.g.,~\cite{Paper-21, Paper-72}), and 3 (e.g.,~\cite{Paper-21, Paper-72}) studies, respectively, addressing these attacks. Other attacks, including TDA, AI model poisoning, phishing, DoS, Cross-Site Scripting, and SQL injection, received minimal attention, were discussed in only one or two studies, and are not illustrated in the figure.

 \begin{figure}[hbt!]
\centering
\includegraphics[width=6cm]{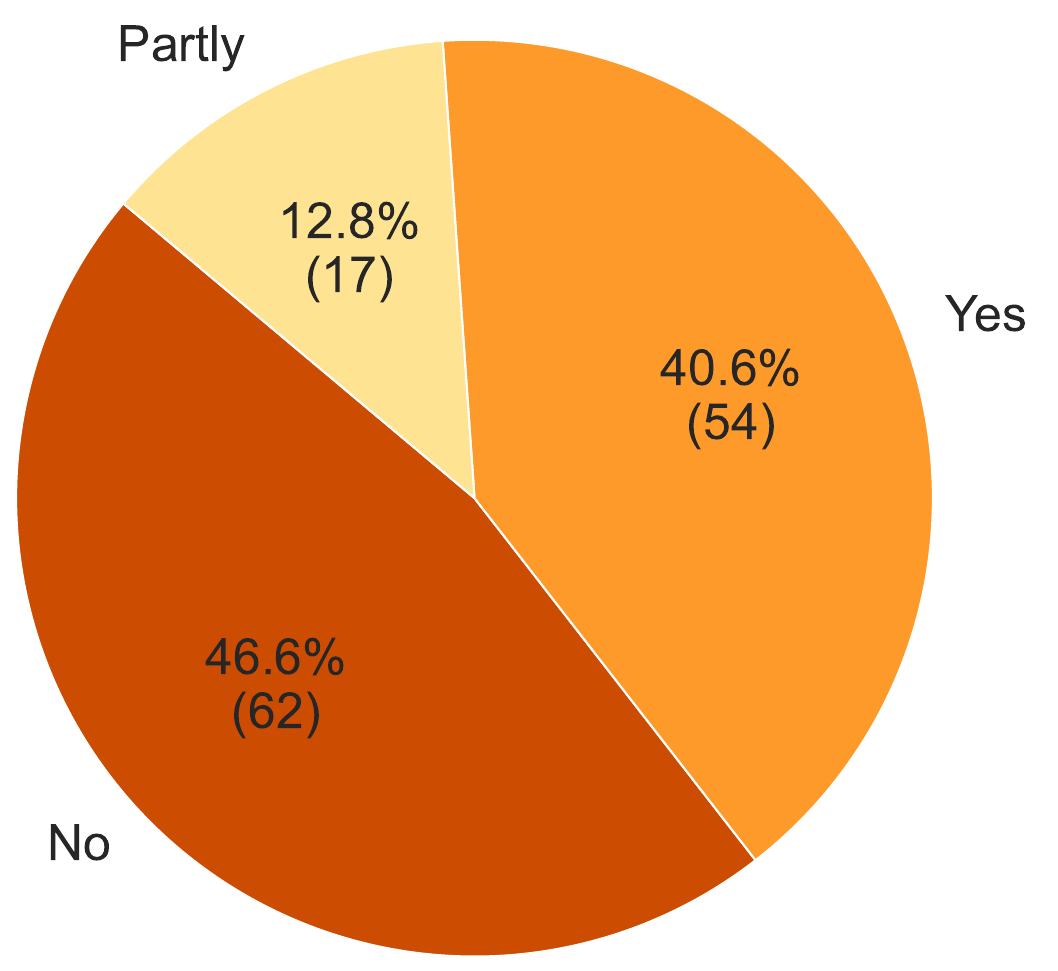}
\caption{Uniqueness of the Addressed Vulnerabilities/Attacks in the Existing Studies.}
\label{fig5k1}
\end{figure}

Addressing specific vulnerabilities and attacks unique to V2G systems, such as load-altering attacks (LAAs), is as crucial as tackling more common vulnerabilities. Solutions designed for typical vulnerabilities may not be directly applicable to these unique threats. Any V2G vulnerability, such as those described in ISO 15118-2~\cite{Paper-18}, or specific attacks like LAAs~\cite{Paper-6, Paper-7}, are deemed unique to V2G systems or their components (e.g., EV and EVCS). If a vulnerability, such as physical access issues related to EVs or EV charging stations \cite{Paper-15, Paper-16}, is also relevant to other systems, we categorise it as a partly unique threat. 

Figure~\ref{fig5k1} illustrates the distinct vulnerabilities and attacks covered in existing research. A significant majority - 82 of 133 studies - focused on common vulnerabilities such as WAu and IC~\cite{Paper-5}, along with related attacks such as malware injection~\cite{Paper-5} and eavesdropping~\cite{Paper-14}. In contrast, only 54 studies (e.g.,~\cite{Paper-10, Paper-23}) examined unique vulnerabilities or attacks of V2G. In comparison, 17 studies (e.g.,~\cite{Paper-15, Paper-27}) investigated partly unique vulnerabilities or attacks. For example,~\cite{Paper-23} seeks to minimise price manipulation on the Energy Market and Trading Platforms, while~\cite{Paper-15} focuses on vulnerabilities arising from physical access. Addressing these unique vulnerabilities is essential for the robust security of V2G systems.

\begin{figure}[hbt!]
\centering
\includegraphics[width=8cm]{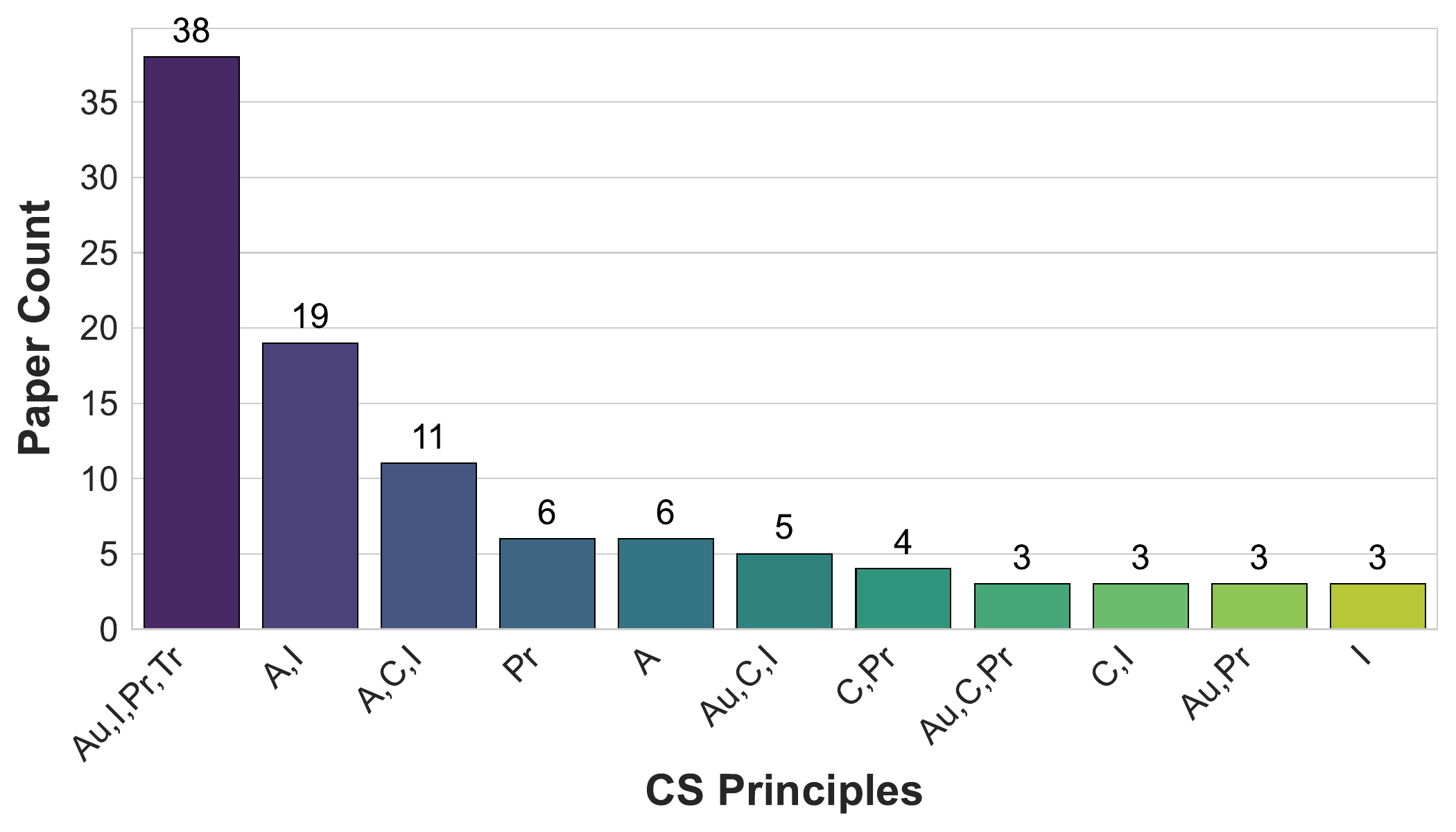}
\caption{Distribution of papers addressing specific combinations of cybersecurity principles.}
\label{fig5k}
\end{figure}

\begin{figure}[hbt!]
\centering
\includegraphics[width=8cm]{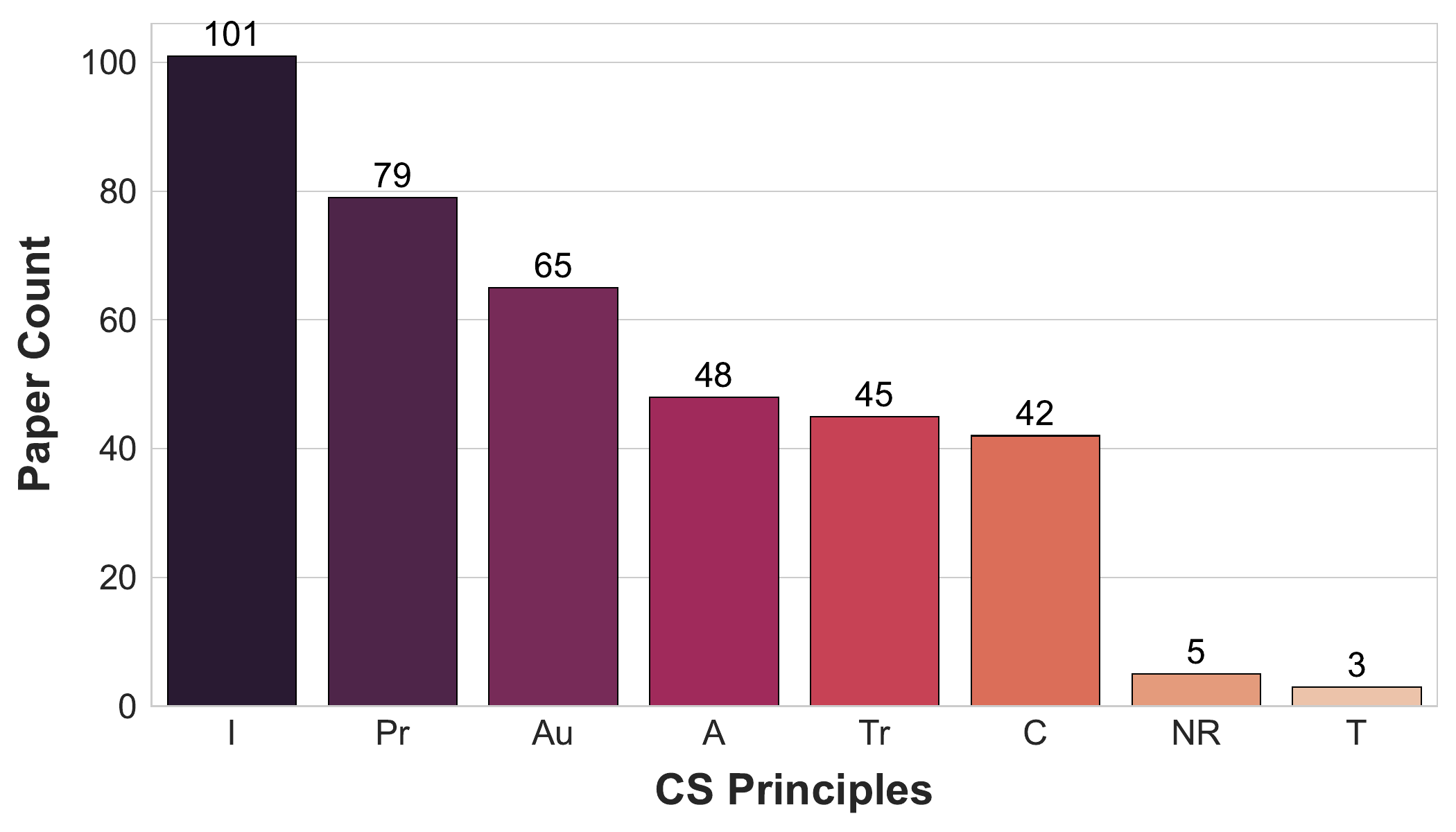}
\caption{Distribution of papers addressing individual cybersecurity principle.}
\label{fig5l}
\end{figure}

\subsection{Cybersecurity Principles}

Cybersecurity principles are fundamental to ensuring the secure operation of V2G systems. The reviewed studies address a wide range of cybersecurity principles, including confidentiality (C), integrity (I), availability (A), authenticity (Au), privacy (Pr), traceability (Tr), trustworthiness (T), and non-repudiation (NR). These principles are critical to protecting V2G systems against various cyber threats and to ensuring the reliability and resilience of the ecosystem. Figure~\ref{fig5k} illustrates the distribution of a specific combination of cybersecurity principles addressed in the reviewed studies, while Figure~\ref{fig5l} provides a detailed breakdown of the sub-principles.

According to Figure~\ref{fig5k}, the combination of Au, I, Pr and Tr is the most common, appearing in 38 studies (e.g.,~\cite{Paper-8, Paper-14, Paper-16, Paper-37}). This combination emphasises the need for secure, confidential and traceable data handling in V2G transactions. Other notable combinations include A and I, addressed in 19 studies (e.g.,~\cite{Paper-3, Paper-4, Paper-18}), and A, C and I, addressed in 11 studies (e.g.,~\cite{Paper-6, Paper-7, Paper-30}). These combinations are essential to ensure that V2G systems remain resilient to cyber threats while maintaining reliable energy/data exchange. Some studies implemented multi-layered security approaches, such as C, I and Pr, along with Au (e.g.,~\cite{Paper-13, Paper-115, Paper-138}), further highlighting the need for comprehensive cybersecurity frameworks.

As shown in Figure~\ref{fig5l}, I and Pr are the most frequently addressed principles, appearing in 101 and 79 studies, respectively. These principles are essential for securing communication channels, protecting sensitive data, and ensuring that information remains unchanged during transmission. For example, studies such as~\cite{Paper-3, Paper-6, Paper-7, Paper-19} emphasise integrity through data validation and error detection mechanisms, while~\cite{Paper-5, Paper-8, Paper-13} propose privacy-preserving techniques, primarily through encryption-based and access control mechanisms. Au was considered in 65 studies (e.g.,~\cite{Paper-12, Paper-14, Paper-19}), often in the context of authentication mechanisms that verify the legitimacy of EVs, users, and energy service providers before granting access to the system. Techniques such as PUF, zero-trust security models and blockchain-backed identity management were commonly used (e.g.,~\cite{Paper-43, Paper-119}). Availability (A) was highlighted in 48 studies (for example,~\cite{Paper-10, Paper-21, Paper-36}), focusing on preventing traditional DoS and distributed DoS attacks that could disrupt V2G operations, and ensuring fault tolerance in the event of network failures. Traceability (Tr) was considered in 45 studies (e.g.,~\cite{Paper-34, Paper-71, Paper-102}), primarily in blockchain-based frameworks where transaction logging and digital signatures help to ensure accountability. Confidentiality (C) was covered in 42 studies (e.g.,~\cite{Paper-15, Paper-19, Paper-109}), focusing on encryption techniques such as Elliptic Curve Cryptography and Homomorphic Encryption to protect sensitive V2G data from interception. However, non-repudiation (NR) and trustworthiness (T) were addressed in only 5 and 3 studies, respectively, indicating a critical research gap in ensuring that entities within a V2G network cannot deny their actions or tamper with security logs.

\begin{table*}[htbp]
\caption{Summary of existing studies based on blockchain in terms of key features.}
\centering
\footnotesize 
\setlength{\tabcolsep}{4pt} 
\begin{adjustbox}{width=\textwidth} 
\begin{tabular}{@{}llllllllll@{}}
\toprule
      &       &           & \multicolumn{2}{c}{Vulnerabilities (Vul) \& Attack} &     &               &             & Methodology &                                           \\
Study & CRMLF & V2G Comp. & Vul & Attacks & U & CS Principles & Technology & Performance & Limitations \\
\midrule
\cite{Paper-8} & P, D & TP, EV & IC, ICS, SV, PA & DM, UA & Yes & Au, I, Pr, Tr & BC, ML & T, EE, Sc & Limited test scenario, privacy issue. \\
\cite{Paper-14} & P & WPT, EV & SV & E, MITM & No & Au, I, Pr, Tr & BC, SC & NA & No real testing, blockchain platform's security issue. \\
\cite{Paper-16} & P, D & EVCS, U, EV & IC, PA & DM, Im, MITM & Partly & Au, I, Pr, Tr & BC, DL & Tr & Lack of blockchain details, scalability issue. \\
\cite{Paper-23} & P & EMTP, SGN, EV & MEVU & DM, FDIA & Yes & A, I, Pr & GT, IOTA & Sc & High computational cost, transaction fees, lacks fairness. \\
\cite{Paper-37} & P & SGN & IC, WAu & DM & Yes & Au, I, Pr, Tr & PBC & Tr & Limited versatility, miner/wallet selection, scalability issues. \\
\cite{Paper-43} & Id, P, D & TP, EVCS, EV & IC, WET & DM, RA, UA & Yes & Au, I, Pr, Tr & BC, ZTA & NA & Centralised trust evaluation, incomprehensive datasets. \\
\cite{Paper-46} & P & EMTP, EVCS, SGN, EV & MEVU & DM & Yes & Au, I, Pr, Tr & BC, QRL & Qs & Focus on energy trading, not cybersecurity. \\
\cite{Paper-47} & P & TP, EVCS, SGN, EV & MEVU & DM & Yes & Au, I, Pr, Tr & BC, CCN & LW & Focus on content delivery, lacks security analysis. \\
\cite{Paper-71} & P & EMTP, EVCS, U, EV & PRV & DM & No & Au, I, Pr, Tr & BC, GT & Ro & Potential scalability challenges, edge computing needed. \\
\cite{Paper-75} & D & SGN, EV & DV, MEVU & DoS, MITM & No & Au, I, Pr, Tr & BC, GT & Ro, Sc & Lacks real-world implementation validation. \\
\cite{Paper-79} & P & EMTP, EVCS, U, EV & DV, WAu & DM, DoS, MITM & Partly & Au, I, Pr, Tr & BC & Sc, Ro & Regulatory challenges in blockchain-V2G integration. \\
\cite{Paper-86} & P & EMTP, SGN, EV & DV, IC, WAu & DM, E, Im, DoS & No & Au, I, Pr, Tr & BC & T & Integration of encryption and blockchain unresolved. \\
\cite{Paper-87} & P & EMTP, EVCS, EV & PRV, WAu & PV & No & Au, I, Pr, Tr & BC, ML & Ro, Sc & Lack comprehensive transaction protection. \\
\cite{Paper-97} & P & PCP, EV & DV, MEVU, SV & DM, DoS & No & Au, I, Pr, Tr & BC & CE, Sc & Focus on charging, ignores discharging capabilities. \\
\cite{Paper-98} & P & V2G & PRV, WAu & MI & No & Au, I, Pr, Tr & BC, GT & NA & Lacks comprehensive security analysis. \\
\cite{Paper-102} & Id, P, R & SGN, EV & PRV & PV & No & Au, I, Pr, Tr & BC, SC & Sc & High computational cost, lacks scalability. \\
\cite{Paper-113} & P & EMTP, EVCS, EV & KMV & DM, UA & No & Au, I, Pr, Tr & BC, GT & EE & Doesn’t address multi-leader scenarios. \\
\cite{Paper-116} & P & EMTP, EV & DV, WAu & FDIA, UA & Yes & Au, I, Pr, Tr & BC, DQN & Sc & No consideration for travel patterns, latency issue. \\
\cite{Paper-117} & P, R & EMTP, EV & PRV & E, Im & Yes & Au, I, Pr, Tr & BC & Tr & Blockchain inefficiencies, lacks auction protocols. \\
\cite{Paper-119} & P, R & V2G & PRV, WAu & MITM & Yes & Au, I, Pr, Tr & BC, SC & Ro & Ideal situation was considered. \\
\cite{Paper-121} & P, Re & EMTP, EVCS, EV & DV, PRV & PV & No & Au, I, Pr, T, Tr & BC, SC & Ro & High computational overhead. \\
\cite{Paper-122} & P & EMTP, EVCS, EV & DV, PRV & DM, DoS, PV & No & Au, I, Pr, Tr & BC & NA & Integration with legacy systems. \\
\cite{Paper-123} & P, R & EMTP, EV & IC & MITM & No & Au, I, Pr, Tr & BC & Ro & Limited cross-platform adaptability. \\
\cite{Paper-126} & Id, P & V2G & IC, PVR & E, RA & No & Au, I, Pr, Tr & BC, SC, ECC & LW & Limited scalability, advanced attack coverage. \\
\cite{Paper-128} & P, R & EVCS, SGN, EV & DV, PVR & DoS, MITM & No & Au, I, Pr, Tr & BC, SC & NA & Real-world adoption challenges. \\
\cite{Paper-129} & Id, P & EVCS, EV & IC, PVR & DoS, MITM & No & Au, I, Pr, Tr & BC, SC & Tr & Privacy-preservation challenges remain. \\
\cite{Paper-144} & P, D, R & EMTP, EVCS, EV & IC, PVR, WAu & DoS & Yes & Au, I, Pr, Tr & BC, SC & Tra & Key Distribution Centre as centralised point of failure. \\
\cite{Paper-149} & P, D, R, Re & EVCS, SGN, EV & DV, VU & DM, P & No & Au, I, Pr, Tr & BC, SC & Rl, Re, Tra & Assumes static grid, hierarchical trust issues. \\
\cite{Paper-151} & P, D, R & SGN, EV & DV & DM, DoS & Yes & Au, I, Pr, Tr & GT, IOTA, SC & NA & Assumes reliable communication, limited V2G focus. \\
\cite{Paper-158} & D, P & EVCS, EV & DV, IC, WAu & MITM, RA & No & Au, I, Pr, Tr & BC, FL & LW, Tra & High memory and speed overhead, scalability concerns. \\
\cite{Paper-159} & P, D, R & EMTP, EVCS, EV & PVR, WAu & MITM, RA & Yes & Au, I, Pr, Tr & BC & NA & No recovery mechanisms, limited WPT integration. \\
\cite{Paper-161} & P, D, R & EMTP, EVCS, SGN, EV & IC, PVR, WAu & Im, MITM, RA & Yes & Au, I, Pr, Tr & BC, O & EE, LL & No recovery mechanisms, limited WPT focus. \\
\cite{Paper-182} & P & EVCS, SGN, EV & IC, PVR & DM, PV & No & Au, I, Pr, Tr & PBC & EE & Scalability in large scale EV and CE issues. \\
\cite{Paper-183} & P & EVCS, SGN, EV & IC, PVR & Im, RA & No & Au, I, Pr, Tr & BC, ECC & NA & Scalability in large scale EV and CE issues. \\
\cite{Paper-185} & P & EVCS, SGN, EV & PVR & PV & No & Au, I, Pr, Tr & BC, DS & NA & Security assumptions lack and limited security analysis. \\
\cite{Paper-187} & P & EMTP, EV & DV, IC, WAu & DM, Im, MITM, PV & Yes & Au, I, Pr, Tr & BC, SDN & LW & Lack of realistic implementation, scalability and throughput. \\
\cite{Paper-188} & P & EMTP, EV & DV, IC & DM, Im & Yes & Au, I, Pr, Tr & BC & Sc, Ro & No dynamic pricing and insecure cooperative EV operations. \\
\cite{Paper-191} & P, D & EVCS, SGN, EV & DV, IC & DM & No & Au, I, NR, Pr & BC, SC, ECC & Sc & Scale issues, user behaviour unrealistic. \\
\cite{Paper-197} & P & EMTP, EVCS, EV & DV, PVR & DM & Yes & Au, I, Pr, Tr & BC, ECC, GT & Tr & Recovery absent; blockchain scale limited. \\
\cite{Paper-198} & P & EVCS, EV & DV, IC & DM & No & Au, I, Pr, Tr & BC, SDN & Re, Sc & Component-specific scale analysis needs expansion. \\
\midrule
\multicolumn{10}{@{}p{26.3cm}}{\scriptsize Acronyms: A (Availability), Au (Authenticity), BC (Blockchain), CCN (Content-Centric Networking), CE (Computational Efficiency), CRMLF (Cybersecurity Risk Management Lifecycle Function), D (Detection), DL (Deep Learning), DM (Data Modification), DoS (Denial of Service), DQN (Deep Q-Network), DV (Data-related Vulnerability), E (Eavesdropping), ECC (Elliptic Curve Cryptography), EE (Energy Efficiency), EMTP (Energy Market and Trading Platforms), EVCS (Electric Vehicle Charging Station), FDIA (False Data Injection Attack), FL (Federated Learning), GT (Game Theory), I (Integrity), IC (Insecure Communication), Id (Identification), ICS (Insecure Communication Protocols), IOTA (Distributed Ledger Technology), KMV (Key Management Vulnerability), LL (Low Latency), LW (Lightweight), MEVU (Malicious EV Users), MI (Malware Injection), MITM (Man-in-the-Middle), ML (Machine Learning), NA (Not Applicable), NR (Non-Repudiation), P (Protection), Pr (Privacy), PA (Physical Access), PBC (Permissioned Blockchain), PCP (Power Control Protocol), PRV (Privacy-related Vulnerability), PV (Privacy Violation), QRL (Quantum Resistant Ledger), Qs (Quantum-safe), R(Response), Re (Recovery), Ro (Robustness), RA (Replay Attack), Re (Resiliency), Rl (Reliability), SC (Smart Contract), Sc (Scalability), SDN (Software Defined Network), SGN (Smart Grid Networks), SV (Software Vulnerability), T (Trustworthiness), Tr (Traceability), Tra (Transparency), UA (Unauthorized Access), V2G (Vehicle-to-Grid), VU (Vulnerable Users), Vul (Vulnerabilities), WAu (Weak Authentication), WET (Weak Encryption Techniques), WPT (Wireless Power Transfer), ZTA (Zero Trust Architecture).}
\\\bottomrule
\end{tabular}
\end{adjustbox}

\label{tab:5.1}
\end{table*}


\begin{table*}[htbp]
\caption{Summary of existing studies based on AI in terms of key features.}
\centering
\footnotesize 
\setlength{\tabcolsep}{4pt} 
\begin{adjustbox}{width=\textwidth} 
\begin{tabular}{@{}llllllllll@{}}
\toprule
      &       &           & \multicolumn{2}{c}{Vulnerabilities (Vul) \& Attack} &     &               &             & Methodology &                                           \\
Study & CRMLF & V2G Comp. & Vul & Attacks & U & CS Principles & Technology & Performance & Limitations \\
\midrule
\cite{Paper-5} & D & EVCS, EV & IC, WAu & MI & No & C, Pr & DL & XAI & Static charging assumption, unrepresentative dataset. \\
\cite{Paper-7} & D, R & EVCS, U, EV & IC, ICS, SV, PA & LAA & Yes & A, I & DL & Re & Fail in unseen attacks, high false negative rate. \\
\cite{Paper-22} & P & EVCS, EV & SNoAI & DM, FDIA & Yes & A & DRL & Sc & Distributed algorithms limit efficiency, scalability issues. \\
\cite{Paper-27} & D & SGN, EV & PA, WAu & DM, FDIA & Partly & A, I & DL & Ac, Ro & IHH features may not reflect practical situations. \\
\cite{Paper-28} & D & EVCS, EV & ICS, SV & DM, FDIA, MITM & Yes & A, I & DL & Ro & Hard to access settings, security limits attacks, CMAs need one-time access. \\
\cite{Paper-29} & D & EVCS, SGN, EV & IC & DM, FDIA & Partly & A, I & DL & Ac & Node-level attack localisation, mitigation challenges in dynamic networks. \\
\cite{Paper-30} & D, R & EVCS, SGN, EV & ICS & DM, FDIA & Yes & A, I & ML, O & Ac & Single-location datasets limit generalisation and the bilevel model oversimplifies. \\
\cite{Paper-42} & D, R & V2G & IC, PA & DM, FDIA, SA & Partly & A, I & DL, DT & LL & Adversaries may have access to resources. \\
\cite{Paper-45} & P & EVCS, EV & DV & MIA, PV & No & Pr & FL, DP & NA & Only binary classification, no multiclass testing. \\
\cite{Paper-53} & D & EVCS, U, EV & ICS & DM, DoS, MITM & Yes & I & DRL & NA & Synthetic data may not generalise well. \\
\cite{Paper-54} & D & EVCS, EV & MEVU, WAu & DoS & No & A & DL & Ac, Ro & Detection effectiveness reduces with weak spatial correlation. \\
\cite{Paper-60} & D & EVCS, EV & MEVU, WAu & DM, MIA, EA, MoIA & Yes & C, I & DL, DP & Ro & Trade-off between accuracy and robustness needed. \\
\cite{Paper-68} & P & SGN, EV & PRV & Im, PV & Yes & Pr & DL & Sc, CoE & Insecure communications within FRL. \\
\cite{Paper-70} & Id, P, D & EMTP, U, EV & DV & DSA & No & A, I & DRL & Ro & Observation based on limited data/information. \\
\cite{Paper-76} & P & EVCS & DV, WAu & DM, MP & No & A, I & FL & LW & Needs practical validation in real environments. \\
\cite{Paper-99} & Id, D, R & EVCS, EV & HV, ICS, WAu & LAA, DoS, MITM & Yes & A, I & ML & NA & Future research needed for effectiveness. \\
\cite{Paper-105} & P, P, R & EV & ICS & FDIA & Yes & Au, I & ML & NA & Limited real-world resilience focus. \\
\cite{Paper-125} & P, D & EVCS, EV & DV & DM & Yes & A, I & DL & NA & Inconsistent BPNN performance. \\
\cite{Paper-131} & Id, P, D & SGN, EV & IC, PVR & DoS, FDIA & No & A, C, I & DL, O & NA & Challenges in generalising DC micro-grid types. \\
\cite{Paper-133} & Id, D, R & EV & HV, IC & FDIA, MITM, RA & No & A, C, I & DL & EE & Offline training, sensitivity issues, limited real-world discussion. \\
\cite{Paper-167} & Id, D & EVCS, U, EV & MEVU & FDIA & Yes & I & DL & NA & Synthetic dataset and limited analysis. \\
\cite{Paper-180} & P & EVCS, SGN, EV & IC, WAu& Im, MITM, PV, RA & No & Au, I, Pr & ML & LW & DY adversary models are not realistic. \\
\midrule
\multicolumn{10}{@{}p{26.3cm}}{\scriptsize Acronyms: A (Availability), Ac (Accuracy), Au (Authenticity), C (Confidentiality), CoE (Communication Efficiency), CRMLF (Cybersecurity Risk Management Lifecycle Function), D (Detection), DL (Deep Learning), DM (Data Modification), DoS (Denial of Service), DP (Differential Privacy), DRL (Deep Reinforcement Learning), DSA (Double Spending Attack), DT (Digital Twin), E (Eavesdropping), EA (Evasion Attack), EE (Energy Efficiency), EMTP (Energy Market and Trading Platforms), EVCS (Electric Vehicle Charging Station), FL (Federated Learning), GT (Game Theory), I (Integrity), IC (Insecure Communication), Id (Identification), LAA (Load Altering Attack), LL (Low Latency), LW (Lightweight), MI (Malware Injection), MIA (Membership Inference Attack), MITM (Man-in-the-Middle), ML (Machine Learning), MoIA (Model Inversion Attack), MP (Model Poisoning), NA (Not Applicable), O (Optimisation), P (Protection), Pr (Privacy), PA (Physical Access), PLS (Physical Layer Security), PT (Physical Tampering), PUF (Physical Unclonable Function), PV (Privacy Violation), R(Response), Re (Recovery), Ro (Robustness), RA (Replay Attack), Re (Resiliency), SA (Switching Attacks), Sc (Scalability), SGN (Smart Grid Networks), T (Trustworthiness), V2G (Vehicle-to-Grid), Vul (Vulnerabilities), WAu (Weak Authentication), T (Wireless Power Transfer), XAI (Explainable AI).}
\\\bottomrule
\end{tabular}
\end{adjustbox}

\label{tab:5.2}
\end{table*}

\begin{table*}[htbp]
\caption{Summary of existing studies based on traditional cryptography in terms of key characteristics.}
\centering
\footnotesize 
\setlength{\tabcolsep}{4pt} 
\begin{adjustbox}{width=\textwidth} 
\begin{tabular}{@{}llllllllll@{}}
\toprule
      &       &           & \multicolumn{2}{c}{Vulnerabilities (Vul) \& Attack} &     &               &             & Methodology &                                           \\
Study & CRMLF & V2G Comp. & Vul & Attacks & U & CS Principles & Technology & Performance & Limitations \\
\midrule
\cite{Paper-2} & P & SGN & IC, PIS & Im, DM & Yes & C, Pr, Tr & DS, CLS, RS & CE & Unrestricted tracer power, anonymity-traceability imbalance. \\
\cite{Paper-13} & D & EV & WAu & Im & No & Au, C, Pr & TC & NA & Only cryptanalysis of an existing protocol. \\
\cite{Paper-38} & P & U, EVCS, SGN, EV & IC, PA & DM, Im, MITM & No & A, C, I & ECC, LWC & LW & No real testing. \\
\cite{Paper-44} & P & V2G & IC & PV & Partly & C, Pr & RS & CoE & Larger ring sizes increase communication cost. \\
\cite{Paper-90} & P & EVCS, SGN, EV & PRV, IC, WAu & DM, DoS, Im, MITM & No & A, C, I & ECC & LW & No security solutions for other V2G scenarios. \\
\cite{Paper-94} & P & EVCS, SGN, EV & HV, IC, WAu & MITM, RA & No & A, C, I & TC & LW & Future focus on inter-entity communication. \\
\cite{Paper-112} & Id, P & TP, EVCS, EV & PRV, ICS & DM, Im & Yes & C, Au, Pr & TC, O & NA & Balancing privacy and learning challenges. \\
\cite{Paper-118} & P & SGN, EV & PRV, VAI & PV & Yes & C, Pr & HE & NA & Limited scalability, insufficient perturbation methods. \\
\cite{Paper-138} & Id, P, D, R & EVCS, EV & IC, PVR, WAu & Im, MITM, RA & Partly & Au, C, I, Pr & ECC & LW & Trusted authority as single point of failure. \\
\cite{Paper-142} & Id, P, D, R & SGN, EV & IC, PVR, WAu & FDIA, PV & Yes & Au, C, I, Pr & HE, DT & CoE & Trusted utility company as single point of failure. \\
\cite{Paper-147} & P, D, R & EVCS, EV & ICS, WAu & RA & No & Au, C, I & ECC & LW, Ro & Assumes reliable communication, no key compromise recovery. \\
\cite{Paper-152} & Id & EVCS, SGN, EV & DV, IC, PA & DM, DoS, FDIA, MITM & No & A, Au, C, I, Pr & TC & NA & Limited scalability, high protocol dependence. \\
\cite{Paper-154} & P, D, R & EVCS, EV & IC, PVR & MITM, RA & Yes & Au, C, Pr & ECC & LW & Assumes reliable cryptographic hardware, limited evaluation. \\
\cite{Paper-155} & P, D, R & EVCS, EV & IC, PVR & MITM, RA & No & Au, C, NR, Pr & TC & LW, Qs & Limited physical security, ideal secure channel assumption. \\
\cite{Paper-157} & D, P & EVCS, EV & DV, PVR & MITM, RA & Yes & Au, C, Pr & TC & LW & Assumes secure initialisation, no recovery mechanisms. \\
\cite{Paper-160} & Id, P, D, R & EVCS, EV & IC & Im, MITM, RA & Yes & C, I, Pr & LWC & LW & No recovery mechanisms, traceability sacrificed. \\
\cite{Paper-162} & P, D, R, Re & EVCS, EV & DV, IC & DM, Im, MITM, RA & Yes & Au, C, NR, Pr, Tr & TC & Ro, Sc & Limited T focus, potential scalability concerns. \\
\cite{Paper-164} & P, D, R, Re & SGN, EV & IC, PA, WAu & MITM, RA & No & Au, C, I, Pr, Tr & ECC, ML & Ro, Sc & Limited energy-specific focus, T vulnerabilities. \\
\cite{Paper-174} & P & U, EV & DV & PV & No & Pr & ECC & NA & Limited detail and lack of generalisability. \\
\cite{Paper-179} & Id, P & EMTP, SGN, EV & IC, WAu & PV & Partly & NR, Pr & CLS & NA & Communication, scalability and security (may not be Qs). \\
\cite{Paper-184} & P & EVCS, SGN, EV & IC, PVR & DoS, Im, MITM, RA, PV & No & Au, I, Pr & TC & LW, Re & Theoretical analysis and simulation tools for validation. \\
\cite{Paper-190} & P, R & EVCS, EV & PVR, WAu & Im, MITM, PV & Partly & C, Au, Pr, Tr & TC & LW, Sc & Personal EVs only and without real-time testbed. \\
\cite{Paper-193} & P & EVCS, SGN, EV & IC, PVR & PV & Partly & I, Pr & ECC & LW & Limited privacy, not talked about specific attacks, scalability issues. \\
\cite{Paper-195} & P, R, Re & EVCS, U, EV & MEVU & MITM, DSA & Yes & I, NR, Pr, Tr & TC & NA & Partial fraud defence, scale unaddressed. \\
\cite{Paper-196} & P, R & EVCS, EV & PVR & FDIA, MITM, RA & Yes & Au, Pr & ECC & NA & Recovery, detection, scaling poorly addressed. \\
\midrule
\multicolumn{10}{@{}p{26.3cm}}{\scriptsize Acronyms: A (Availability), Ac (Accuracy), Au (Authenticity), C (Confidentiality), CoE (Communication Efficiency), CRMLF (Cybersecurity Risk Management Lifecycle Function), D (Detection), DL (Deep Learning), DM (Data Modification), DoS (Denial of Service), DP (Differential Privacy), DRL (Deep Reinforcement Learning), DSA (Double Spending Attack), DT (Digital Twin), E (Eavesdropping), EE (Energy Efficiency), EMTP (Energy Market and Trading Platforms), EVCS (Electric Vehicle Charging Station), FL (Federated Learning), GT (Game Theory), HV (Hardware-related vulnerabilities), I (Integrity), IC (Insecure Communication), Id (Identification), LL (Low Latency), LW (Lightweight), ML (Machine Learning), MITM (Man-in-the-Middle), NA (Not Applicable), NR (Non-Repudiation), O (Optimisation), P (Protection), Pr (Privacy), PA (Physical Access), PIS(Power-related information share), PLS (Physical Layer Security), PT (Physical Tampering), PUF (Physical Unclonable Function), PV (Privacy Violation), Qs (Quantum-Safe), R(Response), Re (Recovery), Ro (Robustness), RA (Replay Attack), Re (Resiliency), Sc (Scalability), SGN (Smart Grid Networks), T (Trustworthiness), V2G (Vehicle-to-Grid), VAI (Vulnerable AI), Vul (Vulnerabilities), WAu (Weak Authentication), WPT (Wireless Power Transfer), XAI (Explainable AI).}
\\\bottomrule
\end{tabular}
\end{adjustbox}
\label{tab:5.3}
\end{table*}

\subsection{Key Technologies}
\label{5.7}

\begin{figure}[hbt!]
\centering
\includegraphics[width=8cm]{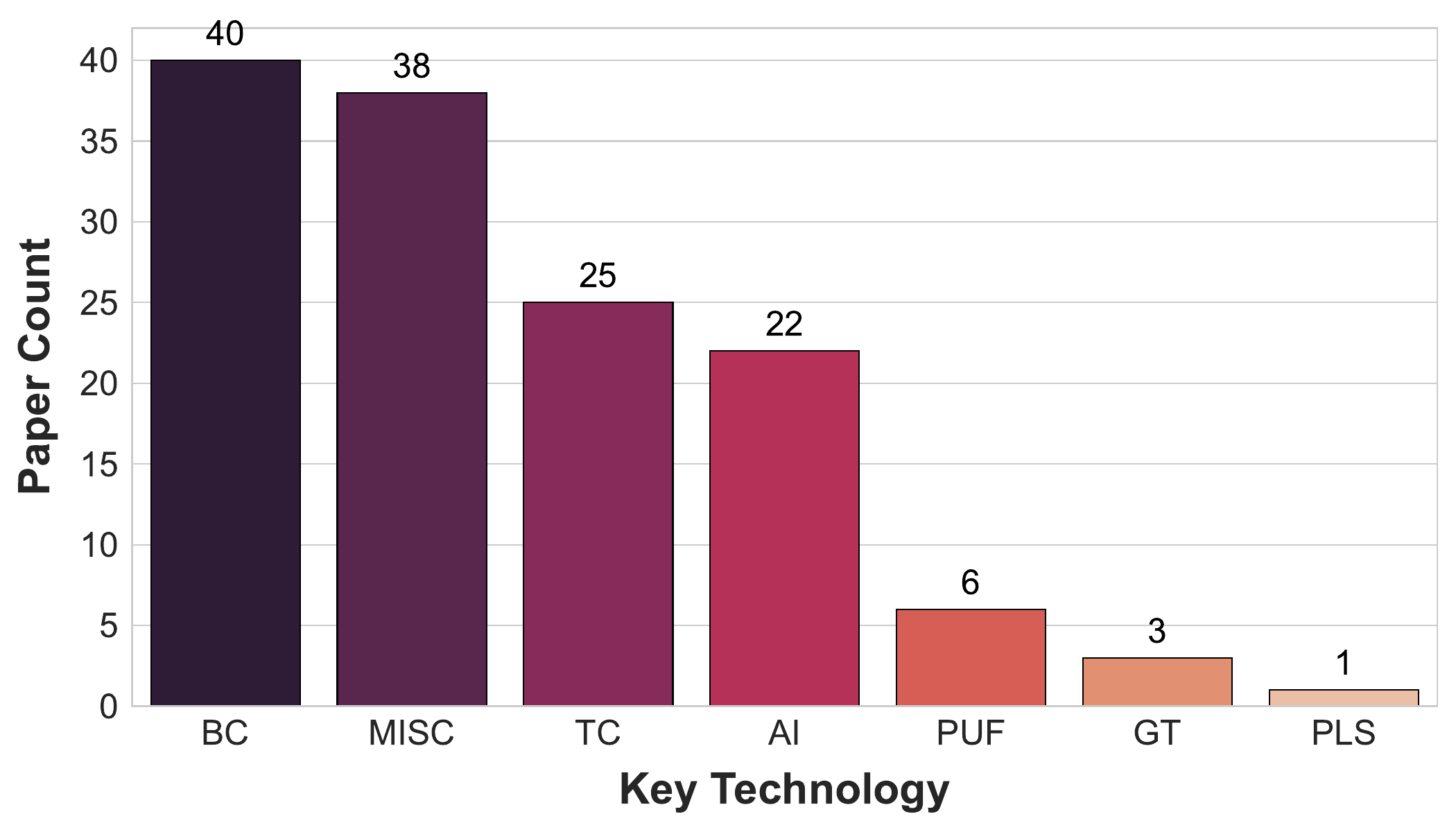}
\caption{Distribution of papers addressing key technology.}
\label{fig5n}
\end{figure}

Cybersecurity in V2G systems has been extensively investigated using a variety of key technologies/methodologies, including AI-driven techniques, cryptographic approaches, blockchain frameworks, and trust models. These technologies address different aspects of security, such as intrusion detection, authentication, secure communication, and risk management. Figure~\ref{fig5n} illustrates the distribution of technologies employed in the reviewed studies, with AI-based approaches (AI), traditional cryptographic techniques (TC), blockchain-based security solutions (BC), and miscellaneous techniques (MISC) being the most prominent. Among these, BC approaches dominate the literature, with 40 studies. A detailed breakdown of major works based on the BC methodology is provided in Table~\ref{tab:5.1}. These studies often integrate blockchain with other technologies, such as smart contracts (SC), machine learning (ML), and zero-trust architecture (ZTA), to enhance transparency, immutability, and decentralised security. In particular, SC-enabled blockchain strategies have received significant attention (e.g.~\cite{Paper-14, Paper-128, Paper-129, Paper-102, Paper-149}). ML techniques have also been used in a supporting role in blockchain-based work, as seen in~\cite{Paper-8, Paper-16}. Furthermore, studies such as~\cite{Paper-71, Paper-75, Paper-151} have combined blockchain with game theory and IOTA to optimise energy trading, while~\cite{Paper-46} introduced a quantum-resistant ledger (QRL) to protect against future quantum computing threats.

AI methodologies have also been widely explored, with federated learning (FL), deep learning (DL), and deep reinforcement learning (DRL) being used in 22 studies (e.g.,~\cite{Paper-5, Paper-7, Paper-22, Paper-42, Paper-68}). A summary of the major studies based on AI methodologies is provided in Table~\ref{tab:5.2}. These approaches are particularly effective in detecting anomalies, predicting cyberattacks, and optimising security protocols. For instance,~\cite{Paper-42} utilised digital twins (DT) and deep learning to simulate and mitigate potential cyber threats, while~\cite{Paper-22} applied DRL to develop adaptive defence mechanisms against dynamic attack scenarios. However, FL and RL approaches have been explored in relatively fewer works, such as~\cite{Paper-45, Paper-53, Paper-70}. TC are the second most widely adopted methodology, appearing in 25 studies (e.g.,~\cite{Paper-2, Paper-13, Paper-38, Paper-90, Paper-152}). These studies primarily focus on secure authentication and data protection mechanisms, with an emphasis on encryption, key management, and lightweight cryptographic solutions (LWC) to secure communication channels between EVs, EVCS, and the grid. For example,~\cite{Paper-38} proposed an authentication scheme based on elliptic curve cryptography (ECC), while~\cite{Paper-152} explored the use of homomorphic encryption (HE) for secure data aggregation in V2G systems. A summary of existing TC studies, highlighting their key characteristics, is presented in Table~\ref{tab:5.3}.

Miscellaneous methodologies (MISC) account for 38 studies (e.g.,~\cite{Paper-3, Paper-18, Paper-19, Paper-36, Paper-39}), encompassing a wide range of approaches, including simulation, optimisation (O) and control theory, as shown in Table~\ref{tab:5.4}. These studies address specific challenges, such as securing hardware components, optimising resource allocation, and simulating attack scenarios. For example,~\cite{Paper-39} used simulation to evaluate the resilience of V2G systems under various attack conditions, while several works (e.g.,~\cite{Paper-3, Paper-21, Paper-33, Paper-134, Paper-178}) studied control theory-based mechanisms to enhance cyber security features in V2G networks. Game theory (GT) is another notable methodology that appears in several studies (e.g.,~\cite {Paper-23, Paper-71, Paper-75, Paper-151}). These studies focus on modelling strategic interactions between attackers and defenders in V2G systems, often in conjunction with blockchain or AI. For example,~\cite{Paper-151} integrated GT with blockchain and smart contracts to design a secure and incentive-compatible energy trading platform. A summary of existing studies based on PUF and game theory, highlighting their key characteristics, is provided in Table~\ref{tab:5.5}. Furthermore, novel approaches such as Digital Twin (DT)-based security models~\cite{Paper-42}, Software-Defined Networking (SDN)-integrated frameworks~\cite{Paper-187, Paper-198}, and Differential Privacy (DP)-enabled security mechanisms~\cite{Paper-60, Paper-66} have received moderate research attention. These emerging methodologies demonstrate the potential for innovative solutions to address evolving cybersecurity challenges in V2G systems. 


\begin{table*}[htbp]
\caption{Summary of existing studies based on miscellaneous technologies (e.g., control theory, optimisation) in terms of key characteristics.}
\centering
\footnotesize 
\setlength{\tabcolsep}{4pt} 
\begin{adjustbox}{width=\textwidth} 
\begin{tabular}{@{}llllllllll@{}}
\toprule
      &       &           & \multicolumn{2}{c}{Vulnerabilities (Vul) \& Attack} &     &               &             & Methodology &                                           \\
Study & CRMLF & V2G Comp. & Vul & Attacks & U & CS Principles & Technology & Performance & Limitations \\
\midrule
\cite{Paper-3} & P, R & EVCS, SGN & IC, PA & DM, FDIA & Yes & I, A & CT & Ro & Missing attack scenarios, EVCS interoperability issue. \\
\cite{Paper-6} & Id & EVCS & IC, ICS, SV, WAu & LAA & Yes & A, I & Botnet, O & NA & Dependency on optimisation algorithms, attacks could be uncertain. \\
\cite{Paper-10} & Id & EVCS & ICS, DS, WAu & E, MITM, LAA & Yes & A & MISC & Sc & Can't identify EVCSs without management systems. \\
\cite{Paper-18} & P & EVCS, EV & IC, WAu & RA & Yes & Au, C, I & TLS & Ro & Incomplete TLS validation, legacy incompatible. \\
\cite{Paper-19} & P & EMTP, EVCS, EV & IC, WAu & DM, Im, MITM, RA & No & C, I & CLS & LW, Tr & Reliant on crypto assumptions, scalability issue. \\
\cite{Paper-21} & Id, R & U, EV & PA, WAu & LAA & Yes & A & CT & NA & Unclear attack traits, hard-to-find equipment, varied standards. \\
\cite{Paper-25} & P & EVCS, SGN & UDSR, UPG & LAA & Yes & A, I & O & Ro & Unproven real-world effectiveness, complex reformulation. \\
\cite{Paper-33} & P, R & SGN, EV & IC & DoS & No & A & CT & Ro & Attack and attacker's capability assumptions could be unrealistic. \\
\cite{Paper-34} & P & TP, EV & WAu & DM, Im, MITM, RA & No & Au, I, Pr, Tr & CLS, BC & Sc & Complex, trust-dependent security raises costs and vulnerabilities. \\
\cite{Paper-36} & P & EV & IC, ICS & TDA & No & A, I & CT, FM & Re & Limited to LTI systems, less applicable to complex systems. \\
\cite{Paper-39} & Id, P, D & V2G & IC, WAu & LAA & No & A, C, I & Simulation & NA & Limited test results. \\
\cite{Paper-40} & P & U, EVCS, SGN, EV & IC, WAu & DM, Im, MITM, RA & No & A, C, I & MISC & LW, Ro & Scalability in large untested V2G networks. \\
\cite{Paper-61} & D & EVCS, SGN, EV & DV, SV & FDIA & Yes & A, I & O & NA & Reliance on communication models and data availability issues. \\
\cite{Paper-64} & Id, D, R & EV & SV & DoS, LAA & Yes & A, C, I & MISC & Sc & Limited app representation in the ecosystem. \\
\cite{Paper-66} & P & SGN & PRV & Im, PV & No & Pr & DP & LW & Scalability issues with high latency. \\
\cite{Paper-72} & R & SGN, EV & DV & LAA & No & A, I, T & CT & Ro & Limited number of EVs used for testing. \\
\cite{Paper-100} & Id, R & EVCS, SGN & DV, MEVU & LAA, DoS & Partly & A, I & SMP & Ro & Focus on individual EVCS, not interconnected networks. \\
\cite{Paper-106} & Id, P, R & EVCS, U, EV & DV, IC & DM & Yes & I, Pr & O & NA & Very ideal scenario considered. \\
\cite{Paper-107} & Id, P, D & EVCS, EV & DV, IC & FDIA & Yes & A, Au, Pr & O & NA & Limited cybersecurity analysis, data restrictions. \\
\cite{Paper-109} & D, R, Re & SGN, EV & CV, HV & DoS & No & C, I, A & MISC & LW & Ineffective against advanced attackers. \\
\cite{Paper-110} & Id, P, D & EVCS & IC, HV & DM, MITM & Yes & C, I, A, Au & TM & NA & False positives reduce accuracy. \\
\cite{Paper-114} & Id, D & EVCS & WAu & XSS, SQLi & No & I, Pr, Tr & MISC & NA & Manual effort required. \\
\cite{Paper-115} & P & EV & PRV & PV & No & C, I, Pr & MISC & NA & Noise reduces accuracy, limited EV personalization. \\
\cite{Paper-120} & P, Id & EV & PRV, WAu & DB, PV & No & Au, Pr & MISC & LL & Limited real-world applications. \\
\cite{Paper-124} & Id, D & EVCS, EV & DV, PA & LAA, SA & No & Au, C, I & MISC & NA & No explicit solution. \\
\cite{Paper-127} & P & EVCS, EV & IC, PVR & MITM, RA & No & A, C & MISC & EE & Limited scalability. \\
\cite{Paper-134} & Id, D, R & EVCS, EV & HV, IC & DoS, FDIA & No & A, C, I & CT & NA & No recovery strategies, limited consideration of nonlinearity. \\
\cite{Paper-135} & Id, P, D, R & SGN, EV & DV, IC, WAu & DM, FDIA & No & A, I & MISC & Ro & No recovery strategies, idealised observer conditions. \\
\cite{Paper-136} & P, D, R & U, SGN, EV & CV, MEVU & LAA, P & Yes & A, I & Simulation & NA & Relies on user awareness, lacks real-time detection. \\
\cite{Paper-145} & P, R & EV & IC & DM, E & Yes & C, I & MISC & Re & Limited practical implementation, no recovery mechanisms. \\
\cite{Paper-146} & P, D, R, Re & SGN, EV & CV, HV & FDIA & No & A, I & MISC & Re & Limited scalability, no recovery mechanisms. \\
\cite{Paper-172} & P & EVCS, EV & IC & Im, RA & No & C, Pr & MISC & LW & Generalisability of the solution and CE. \\
\cite{Paper-173} & Id & EVCS, SGN, EV & DV, PA, WAu & DM, DoS & Yes & A & Simulation & NA & Limited PEVs can make the attack infeasible. \\
\cite{Paper-175} & P, D & EVCS, EV & DV, IC, WAu & PV, DoS, MITM, RA & No & A, Au, Pr & MISC & LW, Ro, Sc & Assumes that EVs are uncompromised during registration. \\
\cite{Paper-177} & P & EMTP, EV & PVR & PV & Yes & Pr & MISC & NA & Limited to only privacy and EMTP. \\
\cite{Paper-178} & D, R & SGN & HV, IC & FDIA & Partly & I & CT & Ro & Simulation and limited system models (68-bus power systems). \\
\cite{Paper-194} & P & EMTP, SGN, EV & PVR & PV & Partly & Pr & MISC & NA & Cloaking area may fail in advance adversaries and scalability issues. \\
\cite{Paper-199} & Id & EVCS & HV & DoS, PT, UA & Partly & A, C, I & TM & NA & Threat modelling only, but no framework used. \\
\midrule
\multicolumn{10}{@{}p{26.3cm}}{\scriptsize Acronyms: A (Availability), Ac (Accuracy), Au (Authenticity), C (Confidentiality), CE (Computational Efficiency), CoE (Communication Efficiency), CRMLF (Cybersecurity Risk Management Lifecycle Function), CT (Control Theory), CV (Control-related Vulnerability), D (Detection), DB (Database Breach), DL (Deep Learning), DM (Data Modification), DoS (Denial of Service), DP (Differential Privacy), DRL (Deep Reinforcement Learning), DT (Digital Twin), E (Eavesdropping), EE (Energy Efficiency), EMTP (Energy Market and Trading Platforms), EVCS (Electric Vehicle Charging Station), FL (Federated Learning), GT (Game Theory), HV (Hardware-related Vulnerabilities), Im (Impersonation), I (Integrity), IC (Insecure Communication), Id (Identification), LAA (Load Altering Attack), LL (Low Latency), LW (Lightweight), MEVU (Metering and EV Usage), MITM (Man-in-the-Middle), ML (Machine Learning), NA (Not Applicable), O (Optimisation), P (Protection), Pr (Privacy), PA (Physical Access), PEVs (Plug-in EVs), PLS (Physical Layer Security), PT (Physical Tampering), PUF (Physical Unclonable Function), PV (Privacy Violation), R(Response), Re (Recovery), Ro (Robustness), RA (Replay Attack), Re (Resiliency), SA (Switching Attack), Sc (Scalability), SGN (Smart Grid Networks), SMP (Security Management Protocol), SQLi (SQL Injection), SV (Software-related Vulnerabilities), T (Trustworthiness), TDA (Time-Delay Attack), TM (Threat Modelling), U (Usability), UA (Unauthorized Access), UDSR (Uncertainty in Demand Side Response), UPG (Uncertainties in Power Generation), V2G (Vehicle-to-Grid), Vul (Vulnerabilities), WAu (Weak Authentication), T (Wireless Power Transfer), XAI (Explainable AI), XSS (Cross-Site Scripting).}
\\\bottomrule
\end{tabular}
\end{adjustbox}
\label{tab:5.4}
\end{table*}

\begin{table*}[htbp]
\caption{Summary of existing studies based on PUF and Game Theory in terms of key characteristics.}
\centering
\footnotesize 
\setlength{\tabcolsep}{4pt} 
\begin{adjustbox}{width=\textwidth} 
\begin{tabular}{@{}llllllllll@{}}
\toprule
      &       &           & \multicolumn{2}{c}{Vulnerabilities (Vul) \& Attack} &     &               &             & Methodology &                                           \\
Study & CRMLF & V2G Comp. & Vul & Attacks & U & CS Principles & Technology & Performance & Limitations \\
\midrule
\cite{Paper-4} & P & T, EVCS, EV & IC, WAu & E, Im, MITM, RA & Yes & Au, C, I & PLS & LL, LW & Artificial noise reliance, scalability issue. \\
\cite{Paper-12} & P & EVCS, SGN, EV & IC, WAu & Im, RA & No & A, C, I, Pr & PUF & LL, LW & No error correction technology for PUFs, scalability issue. \\
\cite{Paper-15} & P & V2G & PA, WAu & E, Im, MITM, RA & Partly & A, C, I & PUF & LW & Protection level unspecified, potentially inadequate. \\
\cite{Paper-17} & P & EVCS, SGN, EV & IC, PA & DM, Im, MITM, RA & Partly & A, I & PUF & LW & DY and CK adversary models are not realistic.\\
\cite{Paper-65} & P & V2G & PA, WAu & DoS, Im, PT, PV & Partly & A, C, I, Pr & PUF & Sc, Ro & PUF may not mitigate advanced physical attacks. \\
\cite{Paper-74} & R & EVCS, EV & IC, PA & DoS & No & A, T & GT & Ro & No emergency recovery or resilience methods. \\
\cite{Paper-80} & P & EVCS, SGN, EV & IC, PA, WAu & DoS, Im, MITM, RA & No & Au, Pr & PUF & LW & Slightly higher communication costs. \\
\cite{Paper-169} & P & EVCS, SGN, EV & PA & Im, MITM, PT, RA & Yes & Au, C, I & PUF & EE, LW & Device-PUF communication must be secure. \\
\midrule
\multicolumn{10}{@{}p{26.3cm}}{\scriptsize Acronyms: A (Availability), Au (Authenticity), C (Confidentiality), CK (Canetti-Krawczyk Adversary Model), CRMLF (Cybersecurity Risk Management Lifecycle Function), D (Detection), DM (Data Modification), DoS (Denial of Service), DY (Dolev-Yao Adversary Model), E (Eavesdropping), EE (Energy Efficiency), EVCS (Electric Vehicle Charging Station), GT (Game Theory), Im (Impersonation), I (Integrity), IC (Insecure Communication), Id (Identification), LL (Low Latency), LW (Lightweight), MITM (Man-in-the-Middle), P (Protection), Pr (Privacy), PA (Physical Access), PLS (Physical Layer Security), PT (Physical Tampering), PUF (Physical Unclonable Function), PV (Privacy Violation), R (Response), Re (Recovery), Ro (Robustness), RA (Replay Attack), Sc (Scalability), SGN (Smart Grid Networks), T (Trustworthiness), V2G (Vehicle-to-Grid), Vul (Vulnerabilities), WAu (Weak Authentication), T (Wireless Power Transfer).}
\\\bottomrule
\end{tabular}
\end{adjustbox}

\label{tab:5.5}
\end{table*}

\subsection{Performance Metrics}

\begin{figure}[hbt!]
\centering
\includegraphics[width=8cm]{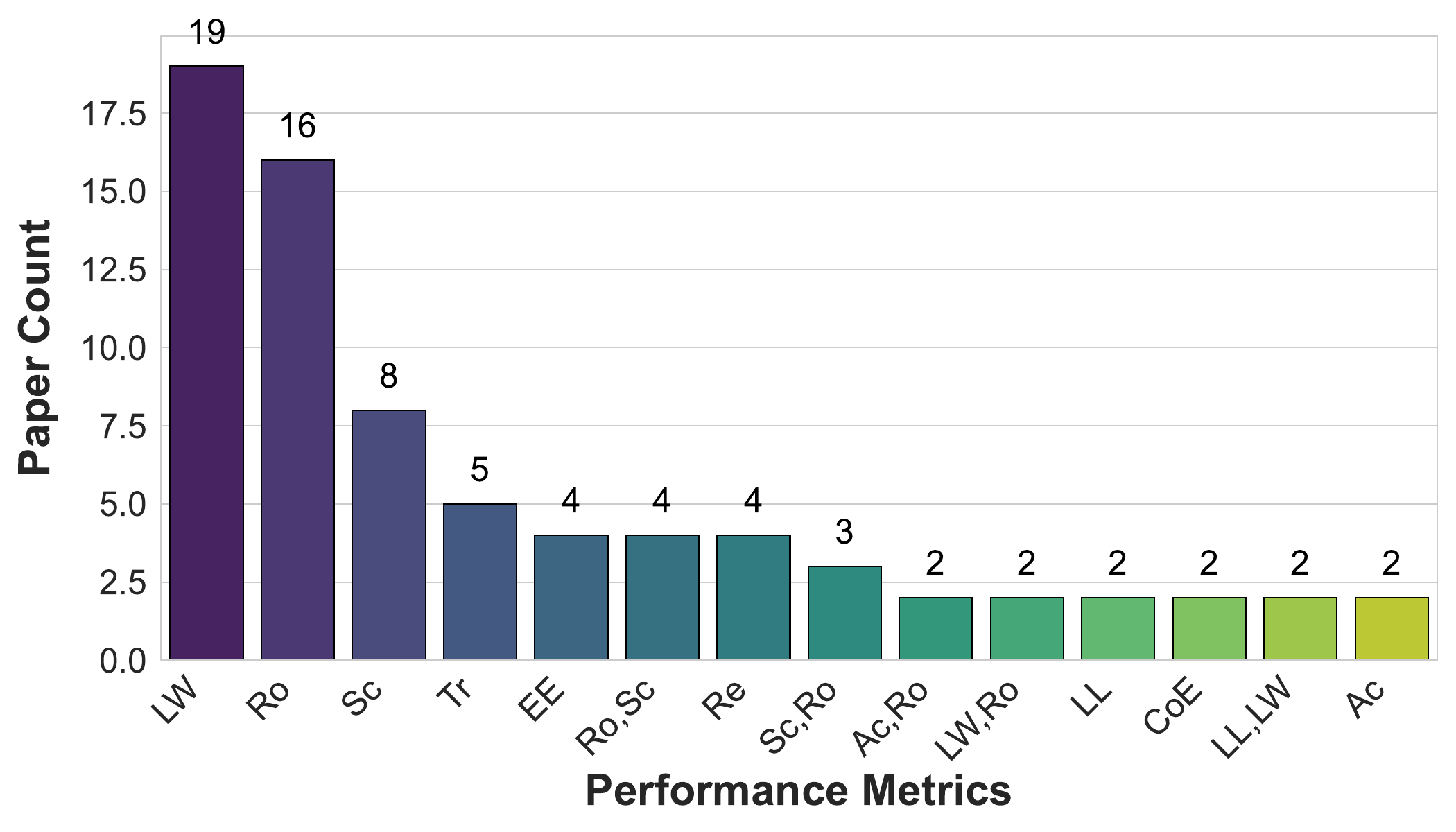}
\caption{Distribution of papers addressing specific combinations of key performance metrics.}
\label{fig5o}
\end{figure}

\begin{figure}[hbt!]
\centering
\includegraphics[width=8cm]{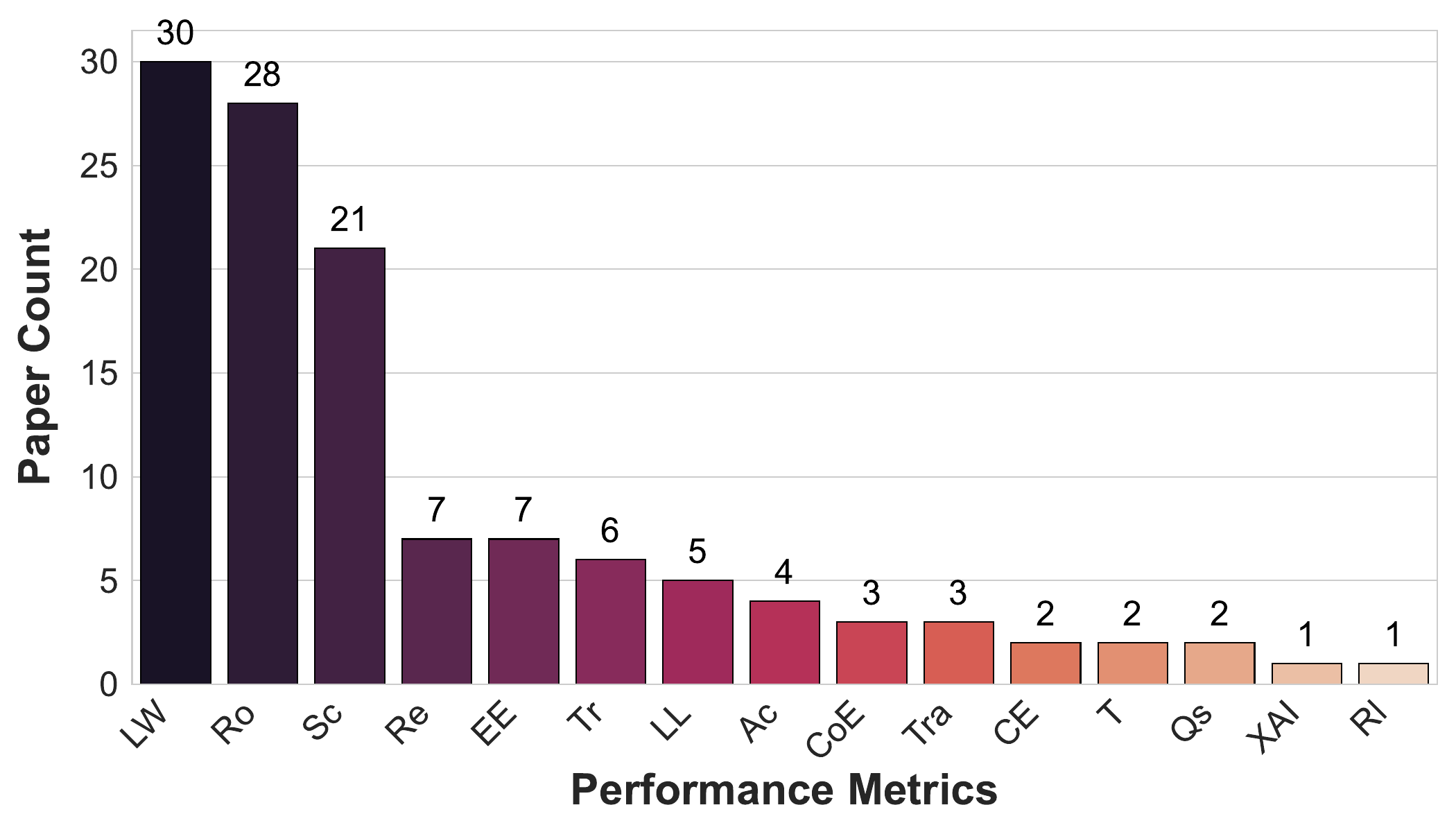}
\caption{Distribution of papers addressing individual key performance metric.}
\label{fig5p}
\end{figure}

Given the importance of V2G systems, evaluating the quality of cybersecurity solutions proposed for these systems is crucial. This evaluation should focus on several key factors, including robustness, scalability, efficiency (energy and computational efficiency), AI assurance, and quantum safety of AI models or encryption methods. Many existing studies tried to address one or two of these quality metrics, and Figure~\ref{fig5o} illustrates the distribution of combinations of these metrics addressed in the reviewed studies. At the same time, Figure~\ref{fig5p} provides a detailed breakdown of the individual metrics.

As shown in Figure~\ref{fig5o} and Tables 3-7, most studies (67 out of 133) have considered at least one key quality metric (e.g., robustness, scalability, efficiency, or AI assurance), while 40 studies did not explicitly address any of these key metrics. The remaining 26 studies evaluated two or three performance or quality metrics; for example, some focused on lightweight (LW) and quantum-safe (Qs) approaches~\cite{Paper-155}, while others considered LW, robustness (R) and scalability (Sc)~\cite{Paper-175}.

In terms of individual metrics, as illustrated in Figure~\ref{fig5p}, many existing studies on V2G cybersecurity have primarily focused on lightweight, robust and scalable solutions, with 30, 28, and 21 studies addressing these aspects, respectively. However, other essential performance metrics remain not explored sufficiently, including reliability, quantum safety of encryption/models, and AI assurance for AI models. For example, only two studies~\cite{Paper-46, Paper-155} have examined the quantum safety of their solutions, focusing specifically on quantum reinforcement learning (QRL)~\cite{Paper-46} and a quantum-safe encryption technique~\cite{Paper-155}. Furthermore, only one study~\cite{Paper-5} has investigated AI assurance in terms of explainability (XAI), while another study~\cite{Paper-149} addressed reliability. Energy efficiency (EE) is considered in 7 out of 133 studies, and resiliency (Re) is addressed in another seven studies. In comparison, other metrics (e.g., traceability and low latency) were discussed between 2 and 5 studies.

\section{Open Research Challenges and Future Directions}
\label{sec6}

Despite advances in the cybersecurity of V2G systems, several unresolved research challenges still need attention. For example, further investigation is required to establish contextualised recovery policies and comprehensive threat modelling at the system level. Furthermore, understanding the connection between the behaviour of EV users and cybersecurity, developing quantum-safe encryption techniques, and ensuring the assurance of AI models remain crucial.

In particular, of the 133 studies conducted, only 22 used AI for cybersecurity solutions. Among these, only one study~\cite{Paper-5} examined AI assurance in terms of explainability. Furthermore, only two studies~\cite{Paper-46, Paper-155} addressed the quantum safety of their AI models or encryption techniques. Although significant progress has been made in V2G cybersecurity solutions, several challenges continue to persist. The following sub-sections outline some of the open research challenges and future directions for the key features of V2G systems.

\subsection{V2G Components}

The cybersecurity of EV and EVCS in V2G systems has been adequately addressed in existing studies. Cyberattack detection in SGN, such as microgrids, has also been explored in ~\cite{Paper-200}. However, none have focused on the behaviour of EV users, despite the fact that humans are often the weakest link in information technology (IT) security systems~\cite {ovelgonne2017understanding}. More research is required to explore the behavioural connection between EV users and cybersecurity attacks and develop potential solutions, including strategies to raise awareness. 

Furthermore, wireless power transfer~\cite{Paper-4}, a potential future charging method, and home or personal charging systems~\cite{Paper-97} are generally more vulnerable (e.g., often connected to insecure home networks) compared to EV charging stations managed by third parties. Unfortunately, these topics have received little attention in the existing literature (only one study~\cite{Paper-97}) and, therefore, need further investigation by the research community.
\subsection{CRLM Functions}

Section~\ref{5.2} indicates significant research focused on detecting and protecting against cyber threats in V2G systems. However, there is a notable lack of studies that address recovery strategies, creating a critical gap in the cybersecurity landscape of these systems. This imbalance is particularly concerning, given the essential role of V2G systems in energy distribution and grid stability. In the event of a cyberattack, rapid and effective recovery is paramount to minimise disruptions, mitigate financial losses, and maintain public trust. The consequences could be severe without robust recovery mechanisms, as evidenced by previous cyber incidents such as the Colonial Pipeline ransomware attack in 2021~\cite{Cybersec19:online}. In addition, the absence of a comprehensive system-wide threat modelling framework to identify potential system-wide threats and risks further deepens the challenge of protecting V2G infrastructure. Therefore, future research should prioritise the development of recovery policies and holistic threat modelling methodologies to strengthen the resilience and security of V2G systems.

\subsection{Vulnerabilities and Attack}

Many vulnerabilities, including insecure communications (IC), weak authentication (WAu), privacy and data violation-related vulnerabilities, are significantly addressed in existing V2G cybersecurity studies. However, issues of physical access, control-related vulnerabilities, AI models, and user behaviour-related attacks in electric vehicle systems have not been adequately addressed and require further research.

Furthermore, existing studies largely overlook vulnerabilities and attacks specific to AI and blockchain technologies. For example, only one~\cite{Paper-76} out of 133 studies has considered AI-specific attacks, such as model poisoning. Considering the importance of AI and blockchain in the cybersecurity of V2G systems, this area requires further investigation to ensure AI assurance and secure blockchain operations in critical systems such as V2G.

\subsection{Key Technologies}

As discussed in~\label{5.7}, blockchain, AI, encryption, control theory, and optimisation are the main technologies used in existing V2G cybersecurity studies. Each of these technologies has specific limitations and open challenges to be addressed as follows:

\begin{itemize}
    \item Blockchain-based V2G cybersecurity research limitations include scalability concerns (e.g.~\cite{Paper-37, Paper-102}), lack of real-world validation (e.g.~\cite{Paper-75}), high computational overhead (e.g.,\cite{Paper-102,Paper-158}, and incomplete security analysis (~\cite{Paper-47, Paper-98}). Future research should focus on scalable architectures, comprehensive security frameworks, efficient consensus mechanisms, and practical implementations while considering dynamic grid conditions and cross-platform integration.
 
    \item Generative AI and transformer architectures have not been used, even though they excel in generating new data sets, representation learning, and temporal data analysis (available in V2G systems), including attack classification or threat analysis~\cite{GneAI-cs25}. Furthermore, existing AI-based V2G cybersecurity faces challenges, including static assumptions~\cite{Paper-5}, limited data sets~\cite{Paper-53, Paper-167}, generalisation issues~\cite{Paper-5, Paper-53}, scalability constraints~\cite{Paper-22, Paper-53}, and the need for external/independent validation~\cite{Paper-75} and robust multiclass detection~\cite{Paper-45}. Future research should focus on developing adaptive, scalable AI-driven cybersecurity frameworks with real-world validation, robust adversary modelling, and multi-class detection to enhance V2G system resilience against evolving cyber threats.
    
    \item The limitations of encryption-based V2G cybersecurity research include a single point of failure in trusted authorities (e.g.,~\cite{Paper-138, Paper-142}), limited scalability (e.g.,~\cite{Paper-152, Paper-179}), and insufficient real-world validation (e.g.,~\cite{Paper-38}). Future research should focus on distributed trust models, efficient cryptographic protocols, and real-world testing and validation of solutions.
     
    \item Control theory and optimisation-based V2G cybersecurity research face limitations in realistic attack modelling (e.g.,~\cite{Paper-106}), scalability (e.g.,~\cite{Paper-40, Paper-66}), system complexity handling~(e.g.,~\cite{Paper-36, Paper-106}), and practical implementation (e.g.,~\cite{Paper-120}). More research is needed on advanced threat modelling, scalable solutions for large networks, and robust and comprehensive real-world validation.
\end{itemize}

\subsection{Performance Metrics}

Several existing studies on V2G cybersecurity have focused on lightweight solutions (30 of 133), robust solutions (28 of 133), and scalable solutions (21 of 133). However, many other important performance metrics remain inadequately addressed, particularly in areas such as quantum-safe encryption/models and AI assurance for AI models. For example, only two studies~\cite{Paper-46, Paper-155} have considered the quantum safety of their solutions, specifically focusing on quantum reinforcement learning (QRL)~\cite{Paper-46} and a quantum-safe encryption technique~\cite{Paper-155}. Furthermore, only one study~\cite{Paper-5} has explored AI assurance in terms of explainability (XAI). Given the critical importance of AI assurance, including the security of AI models and the quantum safety of AI models and encryption, there is an urgent need for further research in these key areas.

\section{Conclusion}
\label{sec:6}

The cybersecurity of V2G systems is necessary to protect these critical cyber-physical system infrastructures and provide uninterrupted services. The focused V2G cybersecurity studies are diverse in terms of (i) the Cybersecurity Risk Management Lifecycle's (CRML) functions, (ii) vulnerabilities and attack vectors or attacks, and (iii) cybersecurity principles (e.g., CIA triad) they address, (iv) considered elements of V2G, (v) key technologies (e.g., Blockchain, AI) in solving cybersecurity issues, and (vi) the quality of the provided solution (e.g., AI assurance, quantum-safe encryption). To provide a holistic view of these diverse studies, we present a systematic review of the field, especially from the perspective of key features of cybersecurity. 

The comprehensive analysis of 133 studies revealed a series of key insights that highlight both achievements and gaps in this growing field. 

\begin{itemize}
   \item Most studies (103 out of 133) focused on protecting V2G systems against cyber threats, while only seven studies addressed the recovery aspect of CRML functions. 
    \item Existing studies have adequately addressed the security of EV and EVCS in V2G systems, but none have focused on the behaviour of EV users and the cybersecurity of V2G systems consists of all six key elements.
    \item Physical access, control-related vulnerabilities, and user behaviour-related attacks in V2G systems are not addressed significantly and overlook vulnerabilities and attacks specific to AI and blockchain technologies. 
    \item Blockchain, AI, encryption, control theory, and optimisation are the main technologies used, and generative AI and the transformer architecture of AI are not used. Finally,
    \item Quantum safety within encryption and AI models and AI assurance is in a very early stage.
\end{itemize}
 
Based on the findings above, certain research areas must be prioritised to address the identified gaps. Therefore, our future efforts will focus on the following elements:

\begin{itemize}
    \item Investigating the relationship between EV users' behaviour and cybersecurity, particularly concerning private charging points.
    \item Using generative AI techniques, such as variational autoencoders for representation learning and feature extraction, generating synthetic attack data, and employing transformer architecture to analyse temporal data on attacks.
    \item Coordinate AI assurance efforts, including all AI goals like security, transparency, safety, and robustness, to create reliable and trustworthy AI systems suitable for real-world applications in V2G systems.
\end{itemize}

\bibliographystyle{model1-num-names}

\bibliography{ref}


%



\end{document}